\documentclass[11pt,preprint]{aastex6}
\usepackage{amssymb}
\usepackage{epsfig}
\usepackage{natbib}
\usepackage{graphicx}
\usepackage{color}
\usepackage{tabularx}
\usepackage{subfigure}

\newcommand       \mum        {\,{\rm \mu m}}

\newcommand       \GBP       {G_{\rm BP}}
\newcommand       \GRP       {G_{\rm RP}}
\newcommand       \Ks           {{ K_{\rm S}}}

\newcommand       \K             {\,{\rm K}}
\newcommand       \simli        {\,{\sim}}
\newcommand       \magni      {\,{\rm mag}}

\newcommand       \gtsim        {\gtrsim}

\newcommand       \Angstrom     {\,{\rm \AA}}

\newcommand       \Teff         {T_{\rm eff}}
\newcommand{\AV}{A_{\rm V}}
\newcommand{\Av}{A_{\rm V}}
\newcommand{\AB}{A_{\rm B}}
\newcommand{\AKs}{A_{\rm K_S}}
\newcommand{\AH}{A_{\rm H}}
\newcommand{\AJ}{A_{\rm J}}
\newcommand{\AG}{A_{\rm G}}
\newcommand{\ABP}{A_{G_{\rm BP}}}
\newcommand{\ARP}{A_{G_{\rm RP}}}
\newcommand{\MRP}{M_{G_{\rm RP}}}
\newcommand{\RV}{R_{\rm V}}
\newcommand{\Rv}{R_{\rm V}}
\newcommand{\RI}{R_{\rm I}}

\countdef\decade=200
\advance\decade by \year
\countdef\hours=201
\advance\hours by \time
\divide\hours by 60
\countdef\mins=202
\advance\mins by \hours
\multiply\mins by 60
\multiply\hours by 100
\countdef\miltime=203
\advance\miltime by \hours
\advance\miltime by \time
\advance\miltime by -\mins

\shorttitle{Optical to Mid-Infrared Extinction Law}
\shortauthors{Wang \& Chen}

\begin{document}

\title{
The Optical to Mid-Infrared Extinction Law Based on the APOGEE, 
{\it Gaia} DR2, Pan-STARRS1, SDSS, APASS, 2MASS and {\it WISE} Surveys
     }

\author{Shu Wang\altaffilmark{1, 2}
             and Xiaodian Chen\altaffilmark{2} 
             }
\altaffiltext{1}{Kavli Institute for Astronomy and Astrophysics,
                 Peking University, 
                 Yi He Yuan Lu 5, Hai Dian District, 
                 Beijing 100871, People's Republic of China;
                 {\sf shuwang@pku.edu.cn}
                 }
\altaffiltext{2}{Key Laboratory for Optical Astronomy, 
                 National Astronomical Observatories, 
                 Chinese Academy of Sciences, 
                 Da Tun Road 20A, Chao Yang District, 
                 Beijing 100101, People's Republic of China;
                {\sf chenxiaodian@nao.cas.cn} 
                 }

\begin{abstract}
A precise interstellar dust extinction law is critically important to interpret observations. There are two indicators of extinction: the color excess ratio (CER) and the relative extinction. Compared to the CER, the wavelength-dependent relative extinction is more challenging to be determined. In this work, we combine spectroscopic, astrometric, and photometric data to derive high-precision CERs and relative extinction from optical to mid-infrared (IR) bands. A group of 61,111 red clump (RC) stars are selected as tracers by stellar parameters from the APOGEE survey. The multiband photometric data are collected from {\it Gaia}, APASS, SDSS, Pan-STARRS1, 2MASS, and {\it WISE} surveys. For the first time, we calibrate the curvature of CERs in determining CERs $E(\lambda - \GRP)/E(\GBP - \GRP)$ from color excess--color excess diagrams. Through elaborate uncertainty analysis, we conclude that the precision of our CERs is significantly improved ($\sigma < 0.015$). With parallaxes from {\it Gaia} DR2, we calculate the relative extinction $\ABP/\ARP$ for 5051 RC stars. By combining the CERs with the $\ABP/\ARP$, the optical--mid-IR extinction $A_\lambda/\ARP$ has been determined in a total of 21 bands. Given no bias toward any specific environment, our extinction law represents the average extinction law with the total-to-selective extinction ratio $\Rv=3.16\pm0.15$. Our observed extinction law supports an adjustment in parameters of the CCM $\Rv=3.1$ curve, together with the near-IR power-law index $\alpha=2.07\pm0.03$. The relative extinction values of {\it HST} and {\it JWST} near-IR bandpasses are predicted in 2.5\% precision. As the observed reddening/extinction tracks are curved, the curvature correction needs to be considered when applying extinction correction. 
\end{abstract}


\section{Introduction}

\subsection{The Optical Extinction}
In ultraviolet (UV)/optical bands of $\lambda < 0.9\mum$, the wavelength-dependent extinction law, $A_\lambda/\AV$, is known to vary from one sightline to another. 
This variation is mainly caused by the change in dust size distributions in different environments. 
The various extinction curves can be approximated by a one-parameter family of curves characterized by the ratio of the total extinction to the selective extinction $\Rv=\AV/E(B-V)=\AV/(\AB-\AV)$ (Cardelli et al.\ 1989, hereafter CCM). 
Theoretically, the extinction produced by Rayleigh scattering of small grains would have a steep curve with $\RV\sim1.2$, while the extinction produced by very large grains would have a flat curve with $\RV \rightarrow \infty$ (Draine 2003). 
Observationally, the $\Rv$ value can be as small as $R_V \sim 2$ in some diffuse sight lines (Fitzpatrick 1999; Wang et al.\ 2017), or as large as $R_V \sim 6$ in dense molecular clouds (Mathis 1990; Fitzpatrick 1999). 
Sight lines toward the Galactic diffuse interstellar medium (ISM) have an average value of $\RV \approx 3.1$ (CCM; Draine 2003; Schlafly \& Finkbeiner 2011). 

Both the color excess ratio (CER) $E(\lambda - \lambda_1)/E(\lambda_2 - \lambda_1)$ and the relative extinction $A_{\lambda}/A_{\lambda_1}$ are indicators of the extinction law. 
Based on observations and the intrinsic color indices of the targets, 
the color excess (CE) $E(\lambda - \lambda_1)$ and the CER $E(\lambda - \lambda_1)/E(\lambda_2 - \lambda_1)$ can be derived.  
However, the calculation of the wavelength-dependent interstellar extinction law $A_{\lambda}/A_{\lambda_1}$ is more challenging. 
It requires an independent determination of the extinction or the distance to the target. 

As a result, many measurements in the literature used a fixed total-to-selective extinction ratio in the optical bands (e.g., $\RV$, $\RI$), or a fixed relative extinction in the near-infrared (NIR) bands (e.g., $\AJ/\AKs$, $\AH/\AKs$) to convert reddenings into extinction law.
For example, Rieke \& Lebofsky (1985) calculated the extinction law for sight lines toward the Galactic center by assuming $\Rv=3.09$.  
Later, Nataf et al.\ (2013) derived the extinction law to the Galactic bulge using a series of fixed total-to-selective extinction ratios $\RI$. 
CCM computed $\Av$ and $\Rv$ by adopting a standard curve given by Rieke \& Lebofsky (1985). 
Indebetouw et al.\ (2005) fitted the red clump (RC) locus in NIR color-magnitude diagrams (CMDs) to directly extract $\AH/\AKs$ by assuming a smooth, homogeneous dust distribution representing an invariant extinction law. However, this assumption is not generally applicable as pointed out by Zasowski et al.\ (2009). 
Gao et al.\ (2009) and Zasowski et al.\ (2009) adopted a fixed NIR extinction value $\AH/\AKs$
to convert CERs into extinction law $A_\lambda/\AKs$. 
The extinction law determined by this method is affected by the systematic uncertainty, which is introduced by the adopted total-to-selective extinction ratio or the NIR relative extinction value.

The distance information provides an opportunity to independently determine the interstellar extinction, $A_{\lambda}/A_{\lambda_1}$. 
However, only some special lines of sight, such as the Galactic center, clusters, and the accurate distances to the objects, can be derived (Fritz et al.\ 2011; Damineli et al.\ 2016; Hosek et al.\ 2018; Chen et al.\ 2018). 
Nishiyama et al.\ (2006, 2009) derived the relative extinction toward the Galactic center by assuming that all RC stars are at the same distance. 
Fitzpatrick \& Massa (2007) determined interstellar extinction curves by analyzing of 328 Galactic stars with known distances. 
Chen et al.\ (2018, hereafter Chen18) adopted classical Cepheids as a diagnostic tool to estimate the relative extinction toward the Galactic center. 
The uncertainties of the extinction law in these measurements are dominated by the accuracy of distance.

\subsection{The Near-infrared Extinction}

The NIR extinction, in the wavelength range of $0.9\mum < \lambda < 3\mum$, follows a power law $A_{\lambda}\propto{\lambda^{-\alpha}}$. 
In the previous century, the index $\alpha$ was estimated to span a small range of $\sim$ 1.6 -- 1.8 (Mathis 1990; Draine 2003, and references therein). 
However, in recent years, with the wealth of deep NIR data, 
much higher power-law indices have been reported, such as $\alpha=$ 1.95 (Wang \& Jiang 2014), 1.99 (Nishiyama et al.\ 2006), 2.05 (Chen18), 2.10 (Wang et al.\ 2013), 2.11 (Fritz et al.\ 2011), 2.14 (Stead \& Hoare 2009), 2.26 (Zasowski et al.\ 2009), 2.30 (Nogueras-Lara et al.\ 2018), and  2.34 (Naoi et al.\ 2007). 
In the lines of sight toward the Galactic center, the index $\alpha$ can even reach 2.5--2.6 (e.g., Gosling et al.\ 2009; Hosek et al.\ 2018). 
Based on these $\alpha$, the corresponding relative extinction $\AJ/\AKs$ ranges from 2.5 to 3.5 (also see Matsunaga et al.\ 2018). 
This means that the uncertainties of the index and the relative extinction are as large as 20\%. 

As mentioned by Fritz et al.\ (2011), one possible explanation is that most of the NIR extinction measurements in the previous century were based on nearby stars ($\sim$ 3 kpc) with low extinction ($\AV \lesssim$ 5 mag),  
while the NIR extinction measured in this century is based on high-extinction sources that are located in the Galactic inner disk, bulge, and even the Galactic center. 
Another possible explanation is the prevailing systematic errors in determining the index $\alpha$.   
When converting the NIR CER (e.g., $E(J-H)/E(J-\Ks)$) to the relative extinction $\AJ/\AKs$ based on the empirical formula $A_{\lambda}\propto{\lambda^{-\alpha}}$, the selection of filter wavelengths (effective or isophotal) affects the value of $\alpha$ and $\AJ/\AKs$.  
As discussed in detail and summarized in Stead \& Hoare (2009), Fritz et al.\ (2011), and Wang \& Jiang (2014), the different choice of filter wavelengths can lead to a significant discrepancy ($\sim 10\%$) in $\alpha$ and $\AJ/\AKs$. 
More specifically, adopting the effective wavelength will lead to a larger $\alpha$, compared to the adoption of the isophotal wavelength. 
Stead \& Hoare (2009) suggested using the effective wavelength instead of the isophotal wavelength that caused the small index value in the measurement prior to 2005. 
However, Fritz et al.\ (2011) found that using the effective wavelength as presented in Stead \& Hoare (2009) would slightly overestimate the index $\alpha$. 
Therefore, Wang \& Jiang (2014) suggested using CERs to represent the NIR extinction. 

In fact, differences are present in the NIR CERs as well. 
Indebetouw et al. (2005) investigated the IR extinction for two regions in the Galactic plane with different environments and derived a constant NIR CER $E(J-H)/E(H-\Ks)=1.778\pm0.156$. 
This value is corroborated by later measurements: 1.72 (Nishiyama et al.\ 2006), 1.78 (Wang \& Jiang 2014), and 1.87 (Xue et al.\ 2016). 
However, some larger  values are also reported, such as $E(J-H)/E(H-\Ks)=2.08\pm0.03$ (Racca et al.\ 2002), $1.91\pm0.01$ (Naoi et al.\ 2007), $2.09\pm0.13$ (Nishiyama et al.\ 2009), and $1.943\pm0.019$ (Schlafly et al.\ 2016). 
To complicate matters, some studies argued that the NIR CERs are varied.
Naoi et al.\ (2006) investigated the extinction toward the $\rho$ Oph cloud and Chamaeleon cloud. They found that $E(J-H)/E(H-\Ks)$ changes with increasing optical depth. 
Later, Zasowski et al.\ (2009) studied the IR relative extinction by RC stars for contiguous sight lines covering $\sim 150^\circ$ of the Galactic disk. 
They reported that the value of IR CER is a function of the angle from the Galactic center, and this variation trend is more obvious in mid-IR bands than that in NIR bands.
The $E(J-H)/E(H-\Ks)$ ranges from 1.95 to 2.18 around the average value $2.04\pm0.06$. 
However, it is worth pointing out that $E(J-H)/E(H-\Ks)$ does not vary much for $\mid l \mid < 60^\circ$. 
Recently, Wang \& Jiang (2014) investigated the NIR extinction law based on a sample of spectroscopic selected K-type giants. 
They reported that the NIR CERs are universal from diffuse to dense interstellar clouds. 
In the measurements of CERs, there are a number of problems. 
One is that the accuracy of CER depends on the purity of the sample and the accuracy of the intrinsic color index. 
Another is that the slope error of the CE--CE diagram is usually underestimated, especially when the sample size or the extinction is small. 
To summarize, the values of NIR CERs still have about 15\% uncertainty and need to be further investigated.

In practice, the NIR extinction value, such as $\AJ/\AKs$, $\AJ/E(J-K_S)$, is commonly used to correct the interstellar extinction. 
Hence, an independent and accurate measurement of the NIR extinction is desired. 
An effort was made by Fritz et al.\ (2011), who used hydrogen emission lines to explore the 1--19 $\mum$ extinction toward the Galactic center via a distance-independent method. 
They derived the absolute extinction and the corresponding $\alpha=2.11\pm0.06$.  
Another effort was made in recent work by Chen18, 
who used classical Cepheids in the direction of the Galactic center to derive the  extinction law between 1 and 8 $\mum$ based on three different approaches. 
They suggested that the values of $\AJ/\AKs=3.005\pm0.031\pm0.094$, $\AH/\AKs=1.717\pm0.010\pm0.033$, and $\alpha=2.05\pm0.07$ can better describe the extinction in the inner Galactic plane.
However, both works were limited to the Galactic center sight lines. 
In this paper, with the precise parallaxes from {\it Gaia} and accurate stellar parameters from APOGEE, we will study the optical to mid-IR extinction in large scale, not just particular sight lines.

\subsection{This Work}

As mentioned in the previous sections, the CERs and relative extinction results in the literature still have significant differences that cannot be fully explained by the uncertainties. 
The precise photometric data ($\sigma \lesssim 0.01$ mag) are helpful to investigate these differences.  
Schlafly et al.\ (2016) have derived accurate CERs in some optical bands by using data from the Pan-STARRS1 (PS1) survey. 
However, as many PS1 stars are too faint to have reliable photometric magnitudes in the Two Micron All Sky Survey (2MASS), {\it Wide-field Infrared Survey Explorer} ({\it WISE}), and APOGEE surveys, the IR extinction has not been well constrained in that work. 
Besides, due to a lack of distance information for the target sources, they measured the reddening curve rather than the extinction curve.

In this work, we try to solve these problems by combining spectroscopic, astrometric, and photometric data. 
The latest data release of the APOGEE survey, DR14, provides us the opportunity to obtain a large and homogeneous sample of RC stars. 
As distance and extinction are usually degenerate, accurate distance information is critical to derive accurate extinction. 
The {\it Gaia} DR2 provides good trigonometric parallaxes for part of RC candidates. 
We gather photometric data from several of survey projects, including APASS, Sloan Digital Sky Survey (SDSS), Pan-STARRS1, 2MASS, {\it WISE}, and the unprecedentedly accurate photometry data from {\it Gaia} DR2. 
Finally, we use RC stars to reinvestigate the optical to mid-IR extinction law. 
Both the reddening curve and the extinction curve have been determined with high accuracy. 
More importantly, we discuss the potential errors in these results in detail.

The description of data sets and the RC star sample are presented in Section 2. 
The optical to mid-IR CERs and relative extinctions are determined in Section 3. 
In Section 4, we analyze the uncertainties of our extinction law in detail. 
We compare our extinction law with those of previous results in Section 5. 
The estimated {\it Gaia} extinction coefficient and predicted NIR extinction values for bandpasses of the {\it Hubble Space Telescope} Wide Field Camera 3 ({\it HST} WFC3) and the {\it James Webb Space Telescope} ({\it JWST}) NIRCAM are also presented in Section 5. 
We summarize our principal conclusions in Section 6.

\section{Data and Sample} \label{data}
\subsection{Data} 

We collect stellar parameters from the APOGEE survey to construct a sample of RC candidates. 
We gather broadband photometric data from the APASS, SDSS, Pan-STARRS1, 2MASS, and {\it WISE} surveys. 
In addition, distance and photometric information from the {\it Gaia} DR2 catalog is extracted. 
By cross-matching these catalogs, stellar parameter, distance, and 21 bands of photometric data from optical to IR are obtained for each star.   

\subsubsection{APOGEE}

The APOGEE (Apache Point Observatory Galaxy Evolution Experiment) is a large-scale, NIR stellar spectroscopic survey (Eisenstein et al.\ 2011). The high-resolution spectra (R$\simli$22500) provide detailed stellar atmospheric parameters (e.g., effective temperature $\Teff$, surface gravity log $g$, metallicity [M/H]) and chemical abundances. 
The primary stellar targets of APOGEE are red giant branch (RGB) stars and RC stars in the bulge, as well as faint stars (Zasowski et al.\ 2013; Abolfathi et al.\ 2018). 
The latest data release, DR14 (Abolfathi et al.\ 2018), contains all data from SDSS-III (APOGEE-1), as well as 2 yr of data from SDSS-IV (APOGEE-2). 
As all APOGEE data, from the beginning of APOGEE-1, were reduced using the latest data reduction pipeline, the parameters provided in DR14 are slightly different from the previous data release version. 

\subsubsection{{\it Gaia}}

The {\it Gaia} mission ({\it Gaia} Collaboration et al.\ 2016) has released the {\it Gaia} DR2, 
in which more than a billion sources have trigonometric parallaxes, three-band photometry ($G$, $G_{\rm BP}$, $G_{\rm RP}$), and proper motions ({\it Gaia} Collaboration et al.\ 2018). 
The $G$ band covers the whole optical wavelength ranging from 330 to 1050 nm, 
while $\GBP$ band and $\GRP$ band cover the wavelength ranges of 330 -- 680 nm and 630 -- 1050 nm, respectively (Evans et al. 2018). 
The central wavelengths of $G$, $G_{\rm BP}$, and $G_{\rm RP}$ bands are 673, 532, and 797 nm, respectively (Jordi et al.\ 2010). 
Concerning the astrometric content, 
for the sources with five-parameter astrometric data, the median uncertainty of the parallax is $\sim$ 0.04 mas for $G <$ 14 mag sources, 0.1 mas at $G$ = 17 mag, and 0.7 mas at $G$ = 20 mag (Lindegren et al.\ 2018). 
Concerning the photometric content, 
the photometric calibrations can reach a precision as low as 2 mmag on individual measurements, and the systematic effects are present at the 10 mmag level (Evans et al.\ 2018).

\subsubsection{APASS} 

The American Association of Variable Star Observers (AAVSO) Photometric All-Sky Survey (APASS) is conducted in five filters: Landolt $B$ and $V$ and Sloan $g', r'$, and $i'$, probing stars with V-band magnitude range from 7 to 17 mag (Henden \& Munari 2014). 
The latest DR9 catalog covers about 99\% of the sky (Henden et al.\ 2016). 
Munari et al.\ (2014) investigated the accuracy of APASS data and confirmed that the APASS photometry did not show any offsets or trends. 
As we also collect SDSS photometric data, we only adopt the $B$ and $V$ data from the APASS DR9 catalog. 

\subsubsection{SDSS}

The SDSS is both an imaging and a spectroscopic survey (York et al.\ 2000). 
The imaging was performed simultaneously in bandpasses $u, g, r, i$, and $z$ with central wavelengths of about 370, 470, 620, 750, and 890 nm, respectively (Fukugita et al. 1996; Gunn et al. 1998).
We take the photometric data from the latest data release DR14, which is the second data release of the fourth phase of the SDSS (Abolfathi et al.\ 2018).

\subsubsection{Pan-STARRS1}

The Pan-STARRS1 survey (Hodapp et al.\ 2004) images the sky in five broadband filters, $g, r, i, z, y$, covering from 400 nm to 1$\mum$ (Stubbs et al.\ 2010).  
The mean 5$\sigma$ point-source limiting sensitivities in $g, r, i, z$, and $y$ bands are 23.3, 23.2, 23.1, 22.3, and 21.4 mag, respectively (Chambers et al.\ 2016).
The effective wavelengths of these filters are 481, 617, 752, 866, and 962 nm, respectively (Schlafly et al.\ 2012; Tonry et al. 2012). 
The photometric accuracy of the PS1 data has been demonstrated by Schlafly et al.\ (2012) and Magnier et al.\ (2013).

\subsubsection{2MASS} 

2MASS is an NIR whole-sky survey (Skrutskie et al.\ 2006). 
The 2MASS point-source catalog contains photometric measurements in the $J$, $H$, and $K_S$ bands with isophotal wavelengths at 1.24, 1.66, and 2.16$\mum$, respectively. 
As the APOGEE objects are selected from 2MASS, the APOGEE catalog already includes the $J$, $H$, $K_S$ measurements. 

\subsubsection{{\it WISE}} 

The {\it WISE} survey is a mid-IR full-sky survey undertaken in four bands: $W1$, $W2$, $W3$, and $W4$ bands with wavelengths center at 3.35, 4.60, 11.56, and 22.09$\mum$, respectively (Wright et al.\ 2010). 
The {\it WISE} photometric data are taken from the AllWISE source catalog. 
Since few sources in our RC sample has reliable $W4$ magnitudes, we only use $W1$, $W2$, and $W3$ data.

\subsection{The Red Clump Sample}

The RC stars are a group of evolved stars in the core helium burning stage. 
They cover the range of spectral types G8III--K2III with effective temperatures of 4500 K--5300 K (Girardi 2016). 
As the luminosities of RC stars are fairly independent of stellar composition and age, 
they are standard candles and widely used to estimate distances in the Galaxy and the Local Group. 
These stars appear as a narrow strip in the CMD, or a clumping group in the effective temperature ($\Teff$)--surface gravity ($\log g$) diagram. 
Therefore, they can be easily selected with photometric or spectroscopic data and become a useful probe to study the interstellar extinction (Indebetouw et al.\ 2005; Gao et al.\ 2009; Wang et al.\ 2017).

On the basis of the available stellar parameters from the APOGEE DR14 survey, we try to construct a homogeneous RC sample with high purity following these steps. 
First, we only include the sources with spectroscopic quality S/N $>$ 50. 
Besides, we limit the metallicity [M/H] $> - 0.5$ dex to reduce the potential effects on RC absolute magnitude. 
Next, RC candidates are selected based on their clumping in the $\Teff$--$\log g$ diagram with $4550\K \le \Teff \le 5120\K$ and $2.2 \le \log g \le 2.8$. 
After these selections, our RC sample contains 61,111 sources. 
A total of 97\% of these RCs have $\Ks$-band magnitude in the range of 7--12.5 mag (distance less than 6 kpc). 
Note that there are some contaminations, such as secondary RC (SRC) stars and RGB stars, in this RC sample. 
We do not remove them in determining CERs (Section 3.2), while we remove them in calculating relative extinction (Section 3.4). 
By cross-matching this RC sample with photometric catalogs listed in Section 2.1, 
the multiband photometric data for RC stars are obtained.
To guarantee the photometric precision, we select stars that satisfy the following criteria for each photometric catalog: 
\begin{enumerate}
\item For {\it Gaia} data, we select stars with photometric error $\le 0.01$ mag and magnitude $\le18.0$ mag in $G,\GBP$, and $\GRP$ bands
\footnote{After the photometric quality selection, there are 56,364 RC stars with {\it Gaia} three-band data. These are further used to determine CERs (Section 3.2). 
In Section 3.4, we remove SRC stars and RGB stars from the parent RC sample and construct a subsample with 30,431 RC stars. 
These RC stars are further selected by the parallax criteria under consideration of subtle biases (Bailer-Jones et al.\ 2018). After that, only 5051 RC stars remained to determine relative extinction $\ABP/\ARP$.}. 
\item For APASS data, we select stars with photometric error $\le 0.05$ mag in $B$, and $V$ bands.
\item For SDSS data, we select stars with photometric error $\le 0.03$ mag and magnitude $\ge 14.0$ mag in $u_{\rm SDSS}$, $g_{\rm SDSS}$, $r_{\rm SDSS}$, $i_{\rm SDSS}$, and $z_{\rm SDSS}$ bands. 
To remove saturated stars, we adopt the criteria $\mid$SDSS/bandpass magnitude-PS1/bandpass magnitude$\mid \le$ 0.5 mag in SDSS/$g, r, i$, and $z$ bands. 
 \item For PS1 data, we select stars with photometric error $\le 0.02$ mag in $g_{\rm PS1}$, $r_{\rm PS1}$, $i_{\rm PS1}$, $z_{\rm PS1}$, and $y_{\rm PS1}$ bands. 
Since many stars in APOGEE are too bright to have reliable photometric magnitudes in PS1, 
we take bright star limit magnitude 14.5, 15.0, 15.0, 14.0, 13.0 to be the brightward limit and  5$\sigma$ single epoch magnitude 22.0, 21.8, 21.5, 20.9, 19.7 (Chambers et al.\ 2016) to be the faintward limit in $g_{\rm PS1}, r_{\rm PS1}, i_{\rm PS1}, z_{\rm PS1}$, and $y_{\rm PS1}$ bands, respectively. 
\item For 2MASS data, we select stars with photometric error $\le 0.03$ mag and magnitude ranging from 6.0 to 14.0 mag in $J, H$, and $\Ks$ bands. 
\item For {\it WISE} data, we select stars with photometric error $\le 0.03$ mag in $W1, W2$, and $W3$ bands.  
\end{enumerate}

\section{The Optical to Mid-IR Extinction Values}

In this section, we calculate the two indicators of the wavelength-dependent extinction law: the CERs and the relative extinction. 
First, we adopt the color--excess method to determine the CERs $E(\lambda - \GRP)/E(\GBP - \GRP)$. 
Then, we derive the relative extinction value $\ABP/\ARP$ by two methods. 
Moreover, combining the CERs with the $\ABP/\ARP$, the optical to mid-IR band relative extinction values $A_{\lambda}/\ARP$ are determined. 

\subsection{Method}
We treat the sample stars as a whole to obtain the extinction by the color--excess method. 
The reason why we do not calculate the extinction of the individual RC star will be discussed in Section 5.1.
Briefly, this method computes the ratio $k_\lambda$ of two CEs 
$E(\lambda - \lambda_1)$ and $E(\lambda_2 - \lambda_1)$, 
which can be expressed as 
\begin{equation} \label{equ1}
k_\lambda=E(\lambda - \lambda_1)/E(\lambda_2 - \lambda_1) = (A_{\lambda} - A_{\lambda_1})/(A_{\lambda_2} - A_{\lambda_1})~~,
\end{equation}
where $A_{\lambda}$ is the extinction in the $\lambda$ band of interest 
and $A_{\lambda_1}$ and $A_{\lambda_2}$ are extinction in the reference $\lambda_1$ band and the comparison $\lambda_2$ band, respectively. 
Therefore, the relative extinction $A_{\lambda}/A_{\lambda_1}$ can be derived by 
\begin{equation} \label{equ2}
A_{\lambda}/A_{\lambda_1} = 1 + k_\lambda (A_{\lambda_2}/A_{\lambda_1} - 1)~~. 
\end{equation}
This method is widely applied to a group of stars with homogeneous intrinsic color indices, such as RGB stars and RC stars (Indebetouw et al.\ 2005; Flaherty et al.\ 2007; Gao et al.\ 2009; Zasowski et al.\ 2009; Wang et al.\ 2013; Xue et al.\ 2016). 

As seen in Equation~(\ref{equ2}), the calculation of $A_{\lambda}/A_{\lambda_1}$ 
requires the knowledge of $A_{\lambda_2}/A_{\lambda_1}$. 
The NIR extinction values $\AJ/\AKs$ and $\AH/\AKs$ are usually used to convert the CERs into the relative extinction $A_{\lambda}/\AKs$ (Section 1.1). 
For example, the relative extinction can be described as $A_{\lambda}/\AKs = 1 + k_\lambda (\AJ/\AKs - 1)$, where $J$ and $\Ks$ bands are treated as the comparison $\lambda_2$ band and the reference $\lambda_1$ band. 
However, as discussed in Section 1.2, the $\AJ/\AKs$ value has 20\% uncertainty for a couple of reasons. 
Compared to the photometric accuracy of 2MASS bands, the photometries of {\it Gaia} bands are at least a factor of 3 more precise. 
Therefore, we take $\GRP$ as the reference $\lambda_1$ band and $\GBP$ as the comparison $\lambda_2$ band to reduce the uncertainties in the extinction determinations. 
The analysis of CER uncertainties caused by adopting different basis bands will be discussed in Section 4.2.

To summarize, we calculate the CER $E(\lambda - \GRP)/E(\GBP - \GRP)$ and the $\ABP/\ARP$, respectively. 
Then, the corresponding relative extinction $A_{\lambda}/\ARP$ is determined by 
\begin{equation} \label{equ3}
A_{\lambda}/\ARP = 1 + k_\lambda (\ABP/\ARP - 1)~~.
\end{equation}

\subsection{Color Excess Ratios}

\begin{figure}[h!]
\centering
\vspace{-0.0in}
\includegraphics[angle=0,width=6.2in]{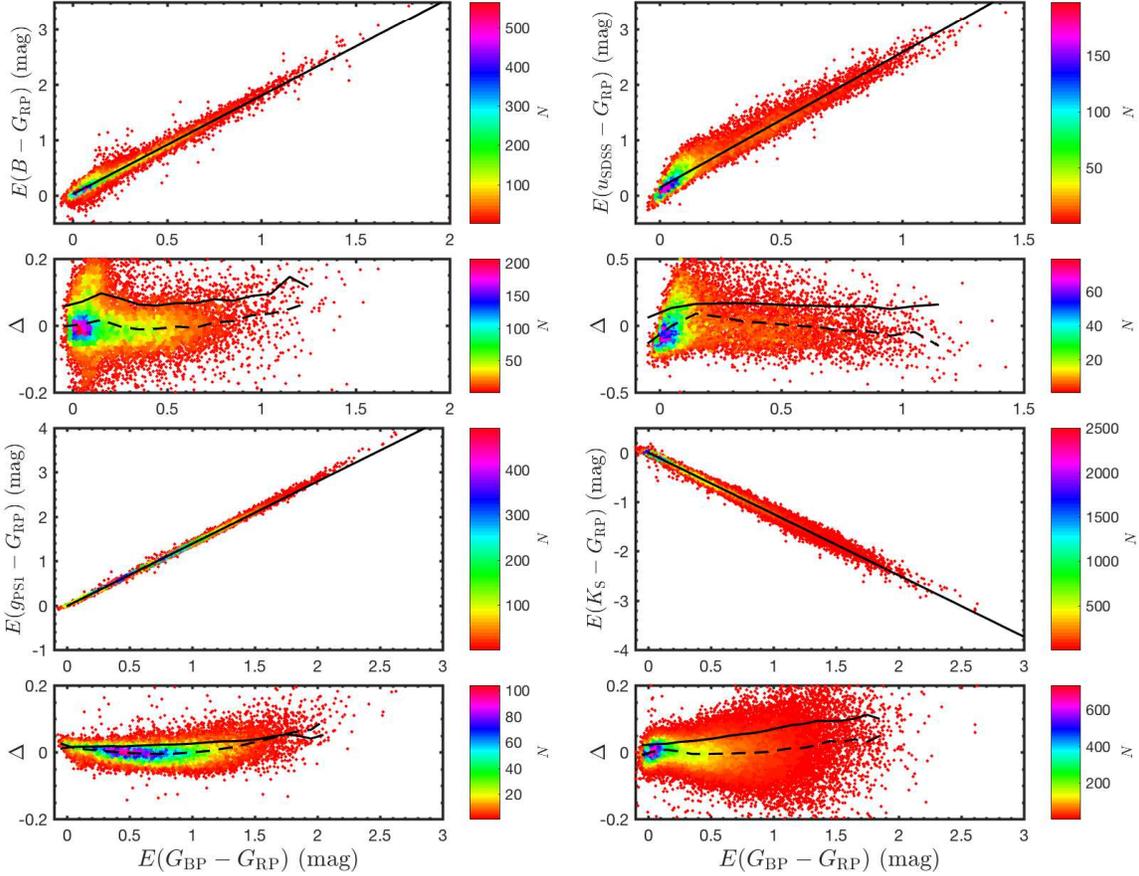}
\vspace{-0.0in}
\caption{\footnotesize
               \label{fig:ccx}
          Measurement of CERs for RC stars 
          in APASS/$B$, SDSS/$u_{\rm SDSS}$, PS1/$g_{\rm PS1}$, and 2MASS/$\Ks$ bands. 
          The color denotes the number density of RC stars. 
          The black lines are the best linear fits. 
          The distributions of the residuals of the fits, $\Delta$, are displayed as well, 
          where solid lines and dashed lines denote the root mean square error (RMSE) 
          and the mean value of residuals, respectively.
               }
\end{figure}

For a group of stars, the CER is the slope $k_\lambda$ of a linear fit to the CE--CE diagram. 
The CE is the difference between the observed color index and the intrinsic color index $E(\lambda_2 - \lambda_1) = (\lambda_2 - \lambda_1)_{\rm obs} - (\lambda_2 - \lambda_1)_{\rm int}$.  
The observed color index $(\lambda_2 - \lambda_1)_{\rm obs}$ 
can be easily obtained from photometric data,  
while the knowledge of intrinsic color index $(\lambda_2 - \lambda_1)_{\rm int}$ 
needs the information of the spectral type (effective temperature) or absolute magnitude. 
Based on the stellar parameters from the APOGEE catalog, we first determine the 
$\Teff$--intrinsic color index relations via the method adopted by Wang \& Jiang (2014). 
The idea of this method is to consider the top 5 \% bluest stars at a given $\Teff$ as the zero reddening star. 
Hence, the observed color index of these bluest stars can represent the intrinsic color index $(\lambda-\GRP)_{\rm int}$ at the given $\Teff$. 
Then, a polynomial fitting is adopted to determine the $\Teff$--$(\lambda-\GRP)_{\rm int}$ relations. 
After subtracting the intrinsic colors, CEs are determined. 
Figure~\ref{fig:ccx} illustrates the linear fit to the CE--CE diagram in four bands, APASS/$B$, SDSS/$u_{\rm SDSS}$, PS1/$g_{\rm PS1}$, and 2MASS/$\Ks$. 
The color shows the number density of RC stars, and the black lines are the best fits. 
The distributions of the residuals of the fits, $\Delta E(\lambda - \GRP)$, are displayed as well.

\begin{table}[h!]
\begin{center}
\caption{\label{tab:int} Coefficients in Equation~(\ref{equ4}) Determining the Intrinsic Color Indices for RC Stars }
\vspace{0.1in}
\begin{tabular}{lcccccccc}
\hline \hline
$(\lambda-\GRP)_{\rm int}$ & $a_2$ & $a_1$ & $a_0$&  $b_2$ & $b_1$ &  $b_0$  &$c_1$ &$c_0$   \\ 
 \hline                                     
{\it GAIA} $\GBP$    & 37.5794&-280.2994 & 523.7778&-0.0001& 0.0313& 0.0035&-0.0140& 0.0343\\
{\it GAIA} $G$       &19.0245 &-141.6585 &264.3243 &-0.0003&-0.0144&-0.0016&0.0082 &-0.0203\\
Johnson $B$    &53.1883 &-399.0859 &750.3023 &0.0877 &0.1005 &0.0078 &-0.0361&0.0900 \\
Johnson $V$    &36.3693 &-271.1687 &506.3027 &0.0280 &-0.0236&-0.0045&0.0150 &-0.0375\\
SDSS $u$       &144.2547&-1080.6096&2026.6185&0.2839 &0.6127 &0.0918 &-0.1351&0.3377 \\
SDSS $g$       &51.6561 &-385.5994 &720.8679 &0.1034 &0.0853 &0.0088 &-0.0149&0.0371 \\
SDSS $r$       &7.6448  &-57.7814  &109.7392 &0.0202 &-0.0172&-0.0050&-0.0073&0.0181 \\
SDSS $i$       &2.8610  &-21.2183  &39.6814  &0.0185 &0.0111 &0.0008 &-0.0095&0.0237 \\
SDSS $z$       &-4.1065 &31.2374   &-59.1555 &0.0246 &0.0298 &0.0043 &-0.0053&0.0132 \\
Pan-STARRS $g$ &45.2888 &-338.3477 &633.1065 &0.0846 &0.0778 &0.0065 &-0.0184&0.0456 \\
Pan-STARRS $r$ &16.6390 &-124.2321 &232.4785 &0.0246 &-0.0061&-0.0023&-0.0091&0.0226 \\
Pan-STARRS $i$ &12.7415 &-94.0923  &174.0668 &0.0244 &0.0266 &0.0025 &-0.0152&0.0376 \\
Pan-STARRS $z$ &-2.0641 &15.9069   &-30.3442 &0.0084 &0.0239 &0.0033 &-0.0078&0.0194 \\
Pan-STARRS $y$ &-15.4901&115.3228  &-214.4557&0.0347 &0.0476 &0.0053 &-0.0090&0.0224 \\
2MASS $J$      &-19.2556&144.7843  &-272.9162&-0.0068&0.0175 &0.0023 &0.0027 &-0.0066\\
2MASS $H$      &-23.3658&177.2650  &-337.3114&0.0373 &0.0764 &0.0067 &0.0045 &-0.0112\\
2MASS $\Ks$    &-27.0262&204.9001  &-389.5475&0.0334 &0.0590 &0.0047 &0.0060 &-0.0150\\
{\it WISE} $W1$      &-27.9958&212.1193  &-403.0451&0.0185 &0.0475 &0.0051 &0.0151 &-0.0377\\
{\it WISE} $W2$      &-36.3077&272.7497  &-513.5114&0.0698 &0.1049 &0.0095 &-0.0287&0.0715 \\
{\it WISE} $W3$      &-14.6125&114.0789  &-223.5171&0.0100 &0.0792 &0.0027 &0.0014 &-0.0035\\

\hline
\end{tabular}
\end{center}
\end{table}

At first glance, the CE--CE  distribution exhibits good linearity, especially in the high-precision $g_{\rm PS1}$ band. 
The dispersions of residuals in $B$ and $u_{\rm SDSS}$ bands are larger than those in $g_{\rm PS1}$ and $\Ks$ bands. 
Moreover, the triangular distributions of residuals are obvious in the low CE part of $B$ and $u_{\rm SDSS}$ bands. 
To reduce the residual, we further analyze the relation among the intrinsic color and metallicity or surface gravity. 
The intrinsic color indices are estimated by stellar parameters, including $\Teff$, [M/H], and $\log g$, which can be expressed as:  
\begin{eqnarray} \label{equ4}
(\lambda - \GRP)_{\rm int}  & = &
       a_2(\log \Teff)^2 + a_1 (\log \Teff) + a_0\nonumber\\
 & &   +   b_2 ({\rm [M/H]})^2 + b_1 ({\rm [M/H]}) + b_0\nonumber\\    
 & &   +   c_1 (\log g) + c_0~~,
\end{eqnarray}
The corresponding coefficients for each color index $(\lambda - \GRP)_{\rm int}$ are listed in Table~\ref{tab:int}. 
We found that the intrinsic colors of the optical bands, such as $B$, $u_{\rm SDSS}, g_{\rm SDSS}, and g_{\rm PS1}$, are correlated with metallicity. 
The intrinsic color index $(u_{\rm SDSS} - \GRP)_{\rm int}$ is even correlated with surface gravity. 
In the other bands, however, the dependence on [M/H] and $\log g$ is moderate or negligible. 
The CEs are then determined by subtracting the intrinsic color indices (Table~\ref{tab:int}) from the observed color indices.

In addition, as seen in Figure~\ref{fig:ccx}, the CER in $g_{\rm PS1}$ band exhibits linear correlations with only $0.05$ root mean square error (RMSE) up to about $E(\GBP-\GRP) = 1.5$ mag. 
After that, the observed extinction track begins to deviate from the linear line, shown as the bowl shape in the residual distribution. 
We further analyze this curvature of CERs in Section 3.3.

\subsection{Curvature of Color Excess Ratios} 

\begin{figure}[h!]
\centering
\vspace{-0.0in}
\includegraphics[angle=0,width=6.5in]{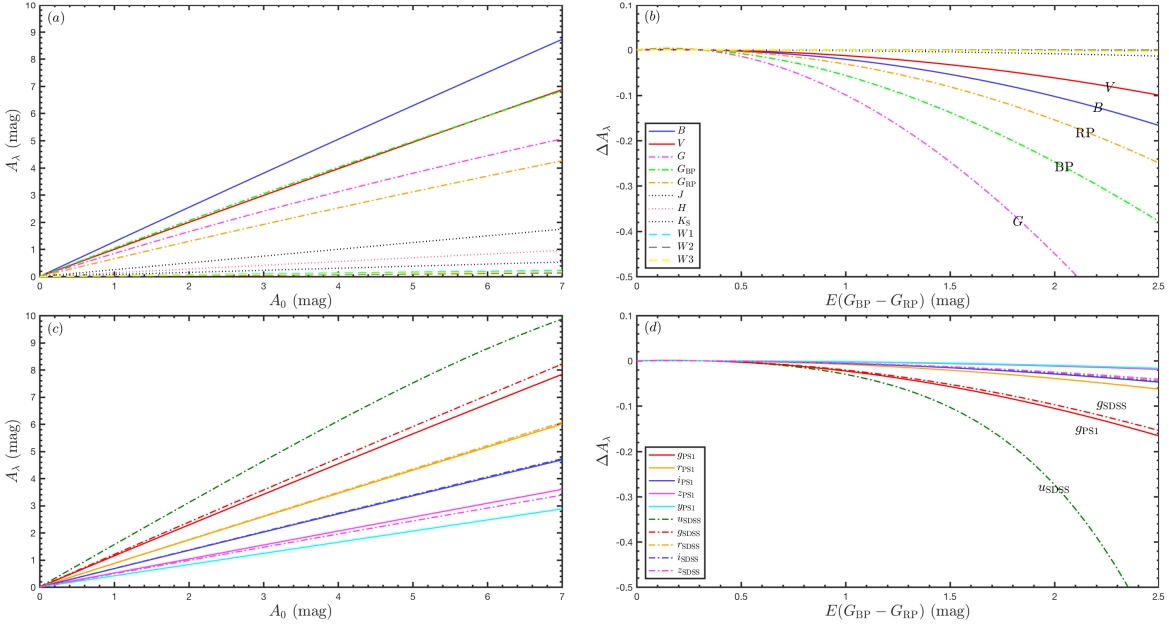}
\vspace{-0.0in}
\caption{\footnotesize
               \label{fig:extcorr}
          Left panels: comparison of 
          the modeled $A_\lambda$ with $A_0$ for each filter. 
          Right panels: difference in extinction $\Delta A_\lambda$ 
          varies with $E(\GBP - \GRP)$. 
          $\Delta A_\lambda$ is the extinction difference between 
          the evolving wavelengths and the static wavelengths based on the RC stellar model.
               }
\end{figure}

The systematic curvature of CERs seen in Figure~\ref{fig:ccx} is due to the assumption of a static wavelength for each filter in fitting the CE--CE diagrams. 
When starlight goes through dust, the peak of spectral energy distribution (the effective wavelength at one band) shifts toward the longer wavelengths gradually, and as a result, the extinction degrades. 
The extinction at a given bandpass filter, namely, evolving filter wavelength extinction, can be expressed as 
\begin{equation}  \label{equ5}
A_\lambda=-2.5 \times \log 
\left (\frac{\int F_\lambda(\lambda)S(\lambda)R(\lambda)d\lambda} 
{\int F_\lambda(\lambda)S(\lambda)d\lambda} \right )~~,   
\end{equation}
where $F_\lambda(\lambda)$ is the intrinsic flux of the stellar spectra and 
$S(\lambda)$ is the filter transmission curve. 
We define $R(\lambda)=10^{-0.4 A_{\lambda,0}}$ as the extinction factor 
and $A_{\lambda,0}$ as the static wavelength extinction. 
According to this formula, the gradually degradation of the extinction is unavoidable, unless the width of the bandpass is infinitely narrow.

In this work, we use {\it Gaia} $\GBP$ and $\GRP$ bands as the basis bands because of their excellent photometric quality, but their broad bandwidth $\Delta \lambda$ would cause significant curvature in the CE--CE diagram. 
To analyze the curvature, we simulate the extinction of each filter band by Equation~(\ref{equ5}). 
We adopt the synthetic stellar spectra $F_\lambda(\lambda)$ for an RC star with $T_{\rm eff}=4800$ K, $\log g=2.5$ and [Fe/H] $= -0.1$ (Lejeune et al.\ 1997), according to the average parameters of the whole RC sample, which are $T_{\rm eff}=4810\pm143$ K, $\log g=2.5\pm0.1$ and [Fe/H] $= -0.1\pm0.2$. 
The spectra are convolved by filter transmission curves of each photometric system.  
The extinction is generated using a CCM $\Rv=3.1$ model extinction curve with $V$-band extinction $A_{V,0}$ from 0 to 20 mag in a step of 0.005 mag.
For comparison, we introduced $A_0$, which denotes the extinction at wavelength 5500 nm with negligible bandwidth.
The left panels of Figure~\ref{fig:extcorr} present the comparison of the modeled $A_\lambda$ with the $A_0$ for each filter, whose slope is the relative extinction $A_\lambda/A_0$. 
The linear lines are obviously bent at $A_0 > 4$ mag in some bands, such as the three {\it Gaia} bands and the $u_{\rm SDSS}$, $g_{\rm SDSS}$, and $g_{\rm PS1}$ bands.

The right panels of Figure~\ref{fig:extcorr} clearly show that the extinction difference between the evolving filter wavelengths and the static wavelengths $\Delta A_\lambda=A_\lambda - A_{\lambda, 0}$ varies with $E(\GBP-\GRP)$. 
If the filter wavelengths did not evolve with the progressive extinction, the extinction tracks, in either photometric system, would have remained at $\Delta A_\lambda= 0$. 
Actually, the $\Delta A_\lambda$ clearly deviates from zero not only in broad {\it Gaia} bands  but also in APASS $B, V$, SDSS $u_{\rm SDSS}, g_{\rm SDSS}$, and PS1 $g_{\rm PS1}$ bands (marked in Figure~\ref{fig:extcorr}). 
Although the deviation of $\Delta A_\lambda$ in the IR bands is less obvious than that in the optical bands, it does exist. 
When $E(\GBP-\GRP) > 9$ mag, corresponding to IR extinction $E(H-\Ks) > 1.2$ from the reddening law of Table~\ref{tab:cc}, the curvature becomes apparent. It can reach to $\Delta E(J-H)=0.15$ mag at $E(H-\Ks)=2.0$ mag (see Fig.\ 2 of Stead \& Hoare 2009).

In the previous literature, the curvature of CERs is not significant in the color--color or CE--CE diagrams, due to the low photometric accuracy or low extinction. 
One exception is Stead \& Hoare (2009), who derived the NIR reddening law by  considering the curvature of NIR CERs caused by filter wavelengths that evolve with the changing spectra of progressively reddened objects.
As shown in Figure~\ref{fig:extcorr}, the curvature of CERs appears at $E(\GBP-\GRP) \sim 0.5$ mag ($\AV \sim 1.2$ mag) for some bands. 
Therefore, the curvature correction requires special attention as the quality of the photometry improves.

\begin{figure}[h!]
\centering
\vspace{0.0in}
\includegraphics[angle=0,width=6.0in]{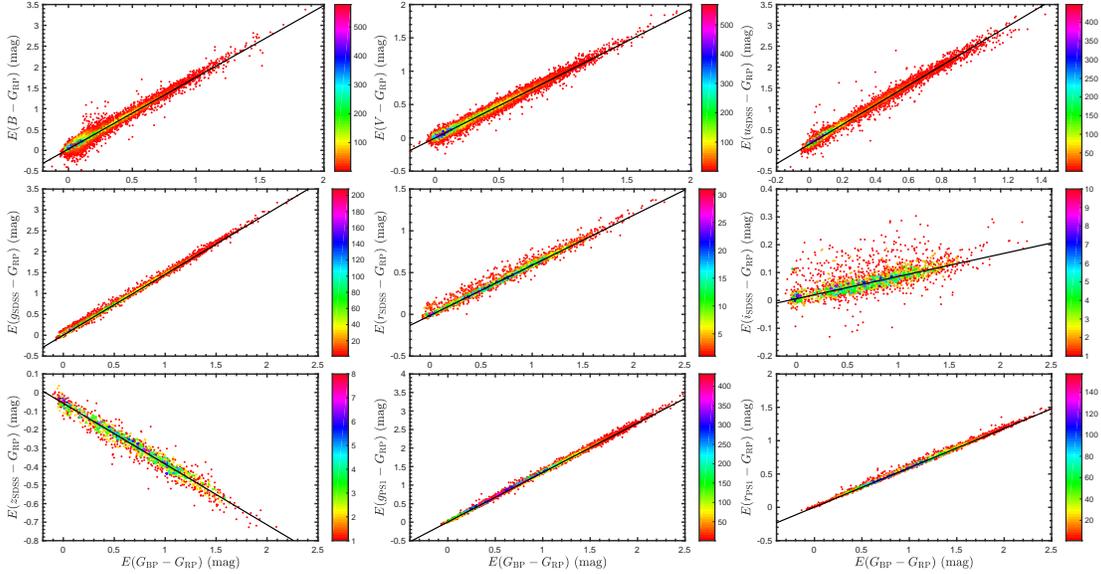}
\vspace{-0.0in}
\caption{\footnotesize
               \label{fig:cc1}
          Color excess--color excess diagrams 
          $E(\GBP - \GRP)$ vs. $E(\lambda - \GRP)$ of RC stars, 
          where $\lambda$ are 
          $B$ and $V$ bands from APASS, 
          $u_{\rm SDSS}$, $g_{\rm SDSS}$, $r_{\rm SDSS}$, 
          $i_{\rm SDSS}$, and $z_{\rm SDSS}$ bands from SDSS, and 
          $g_{\rm PS1}$ and $r_{\rm PS1}$ bands from PS1, respectively, 
          from the top left to the bottom right. 
          The color represents the number density of RC stars.  
          The black lines are the best linear fits to the data, 
          and the slopes (CERs) are listed in Table~\ref{tab:cc}.
               }
\end{figure}

\begin{figure}[h!]
\centering
\vspace{-0.0in}
\includegraphics[angle=0,width=6.0in]{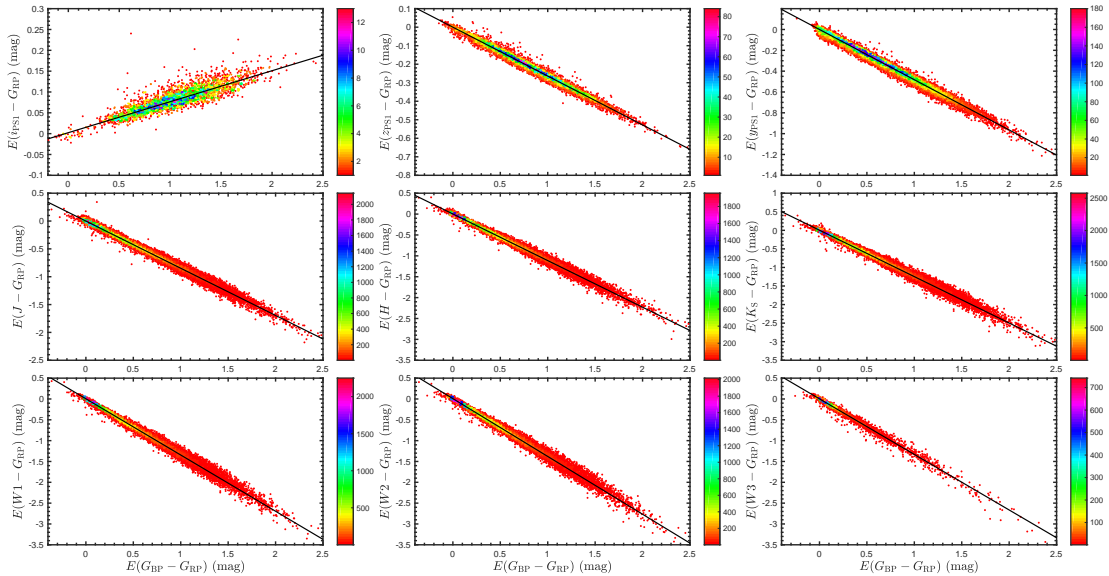}
\vspace{-0.0in}
\caption{\footnotesize
               \label{fig:cc2} 
          Same color excess--color excess diagrams 
          $E(\GBP - \GRP)$ vs. $E(\lambda - \GRP)$   
          as in Figure~\ref{fig:cc1}, but $\lambda$ are 
          $i_{\rm PS1}$, $z_{\rm PS1}$, and $y_{\rm PS1}$ bands from PS1, 
          $J$, $H$, and $\Ks$ bands from 2MASS, and 
          $W1$, $W2$, and $W3$ bands from {\it WISE}, respectively,  
          from the top left to the bottom right.
               }
\end{figure}
\begin{table}[h!]
\begin{center}
\caption{\label{tab:cc} Multiband Color Excess Ratios}
\vspace{0.1in}
\begin{tabular}{lccccc}
\hline \hline
Band ($\lambda$) & $N$   & $E(\lambda-\GRP)/E(\GBP-\GRP)$\tablenotemark{a} & $\sigma$ & (RMSE)$_{\rm max}$ &$(\overline \Delta)_{\rm max}\tablenotemark{b}$\\
 \hline                                                                                    
Johnson $B$      & 15808 & $1.722 \pm0.003\pm0.015$   & 0.052  &  0.079  &  0.022   \\ 
Johnson $V$      & 20487 & $0.965 \pm0.001\pm0.004$   & 0.033  &  0.044  &  0.010   \\ 
SDSS $u$         & 12523 & $2.362 \pm0.003\pm0.009$   & 0.045  &  0.072  &  0.024   \\ 
SDSS $g$         & 6540  & $1.454 \pm0.001\pm0.001$   & 0.023  &  0.034  &  0.022   \\ 
SDSS $r$         & 3546  & $0.601 \pm0.001\pm0.002$   & 0.022  &  0.029  &  0.008   \\ 
SDSS $i$         & 2372  & $0.080 \pm0.002\pm0.001$   & 0.021  &  0.025  &  0.006   \\ 
SDSS $z$         & 1594  & $-0.333\pm0.002\pm0.000$   & 0.021  &  0.027  &  0.003   \\ 
Pan-STARRS $g$   & 26314 & $1.335 \pm0.001\pm0.001$   & 0.019  &  0.033  &  0.008   \\ 
Pan-STARRS $r$   & 9223  & $0.590 \pm0.001\pm0.000$   & 0.012  &  0.020  &  0.007   \\ 
Pan-STARRS $i$   & 2996  & $0.074 \pm0.001\pm0.001$   & 0.014  &  0.017  &  0.004   \\ 
Pan-STARRS $z$   & 9542  & $-0.264\pm0.000\pm0.000$   & 0.012  &  0.014  &  0.010   \\ 
Pan-STARRS $y$   & 21403 & $-0.482\pm0.001\pm0.000$   & 0.021  &  0.029  &  0.007   \\ 
2MASS $J$        & 54914 & $-0.847\pm0.001\pm0.001$   & 0.028  &  0.045  &  0.007   \\ 
2MASS $H$        & 45134 & $-1.116\pm0.001\pm0.002$   & 0.033  &  0.050  &  0.017   \\ 
2MASS $\Ks$      & 55427 & $-1.250\pm0.001\pm0.001$   & 0.034  &  0.054  &  0.012   \\ 
{\it WISE} $W1$        & 42609 & $-1.350\pm0.001\pm0.001$   & 0.031  &  0.053  &  0.017   \\ 
{\it WISE} $W2$        & 42469 & $-1.388\pm0.001\pm0.001$   & 0.034  &  0.057  &  0.007   \\ 
{\it WISE} $W3$        & 8064  & $-1.334\pm0.006\pm0.006$   & 0.026  &  0.042  &  0.008   \\ 
GAIA $G$         & 56364 & $0.461 \pm0.000\pm0.000$   & 0.005  &  0.008  &  0.001   \\ 
   
\hline
\end{tabular}
\tablenotetext{a}{The statistical and systematic errors in the CERs are indicated (in this order).} 
\tablenotetext{b}{The $\sigma$ is the dispersion of the fit. The $\rm (RMSE)_{\rm max}$ and the $(\overline \Delta)_{\rm max}$ are the maxima of the RMSE and the $\overline \Delta$, respectively, where the RMSE and the $\overline \Delta$ are the root mean square error and the mean value of residuals for stars in bins of $\Delta(\GBP - \GRP)= 0.1$ mag, respectively.}
\end{center}
\end{table}

We use the extinction tracks in Figure~\ref{fig:extcorr} (b) and (d) to remove the observed curvature of CERs. 
Figures~\ref{fig:cc1} and~\ref{fig:cc2} are the CE--CE diagrams $E(\lambda-\GBP)$ versus  $E(\GBP - \GRP)$ after the curvature correction, 
where $\lambda$ include 
two APASS bands $B$, and $V$; 
five SDSS bands $u_{\rm SDSS}, g_{\rm SDSS}, r_{\rm SDSS}, i_{\rm SDSS}$, and $z_{\rm SDSS}$; 
five PS1 bands $g_{\rm PS1}, r_{\rm PS1}, i_{\rm PS1}, z_{\rm PS1}$, and $y_{\rm PS1}$, 
three 2MASS bands $J, H$, and $\Ks$;  
and three {\it WISE} bands $W1, W2$, and $W3$. 
The {\it Gaia} $G$-band CE--CE diagram is shown separately in Figure~\ref{fig:Gbprp}. 
The color represents the number density of RC stars. 
The black lines are the best linear fits, and the slopes $k_\lambda$ are CERs. 
For each $\lambda$ band, the number of RC stars that participate in the linear fit, the CERs $E(\lambda-\GRP)/E(\GBP-\GRP)$, and the dispersion of the fit $\sigma$ are tabulated in the second, third, and fourth columns of Table~\ref{tab:cc}, respectively. 
The fitting error of the slope is only $\lesssim$ 0.001, which could not represent the real error of CER. 
Therefore, we discuss the uncertainties of CERs in Section 4.

\subsection{Relative Extinction}\label{A/A}

\begin{figure}[h!]
\centering
\vspace{-0.0in}
\includegraphics[angle=0,width=6.5in]{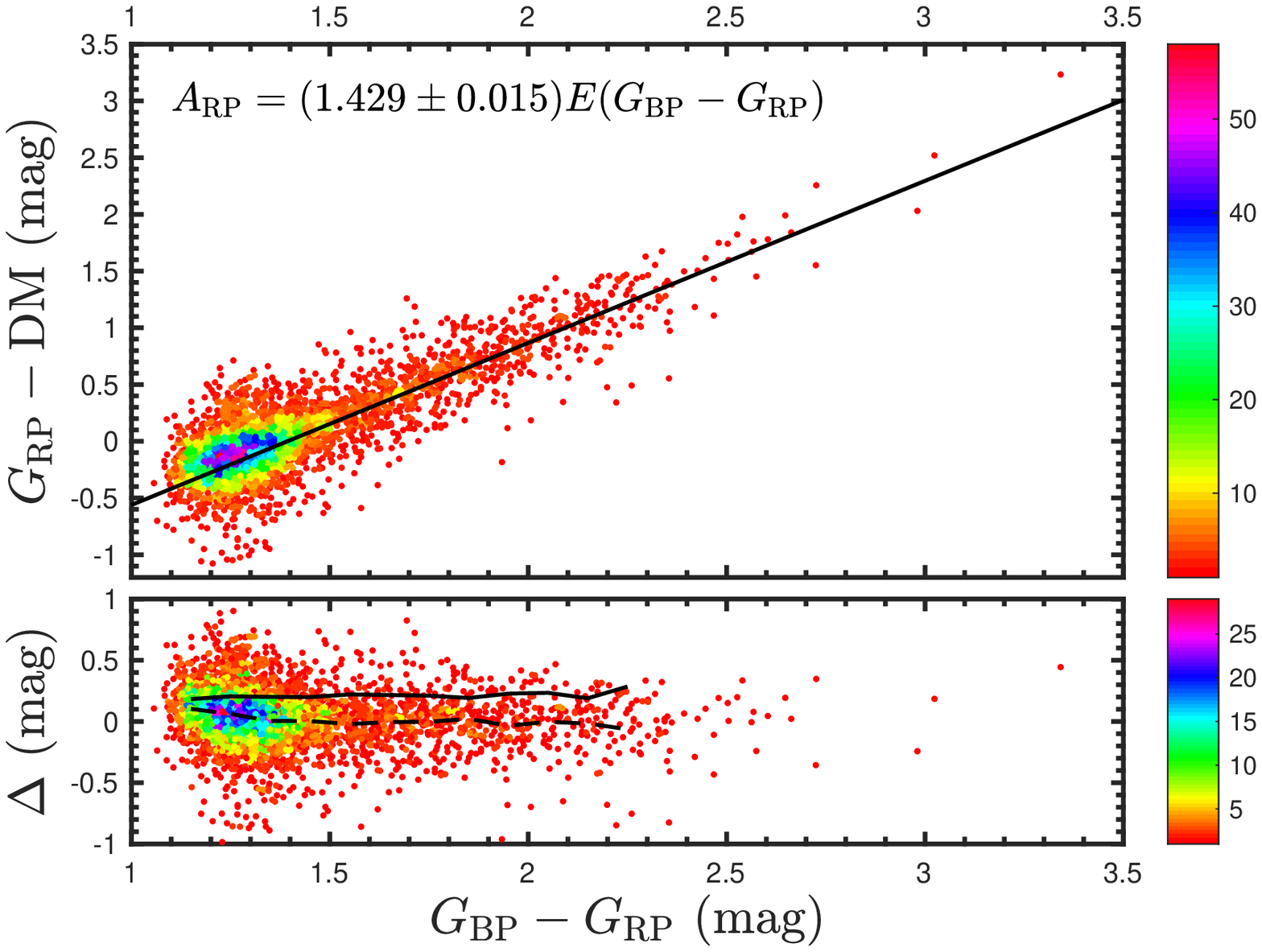}
\vspace{-0.0in}
\caption{\footnotesize
               \label{fig:ABP/ARP}
          Top panel: color excess--extinction diagram for 5051 RC stars, 
          colored by the number density of RC stars. 
          The $x$-axis is the observed color index $\GBP - \GRP$, 
          the $y$-axis is apparent magnitude minus distance modulus 
          $\GRP - {\rm DM} = \ARP + \MRP$, 
          and the black line is a linear fit, which denotes the extinction direction. 
          Bottom panel: distribution of residuals $\Delta(\GBP - {\rm DM})$. 
          The solid line and the dashed line are the RMSE and the mean value of residuals  
          for stars in bins of $\Delta(\GBP - \GRP)= 0.1$ mag, respectively.  
               }
\end{figure}

\begin{table}[h!]
\begin{center}
\caption{\label{tab:ext} Multiband Relative Extinction Values}
\small
\vspace{0.1in}
\begin{tabular}{lccccc}
\hline \hline                                                      
Band ($\lambda$) & $\lambda_{\rm eff, 0}$ ($\mum$) 
& $A_\lambda/\ARP$&  $A_\lambda/\ARP$(from Chen18) 
&  $A_\lambda/A_V$  &  $A_\lambda/E(\GBP-\GRP)$ \\ 
 \hline
{\it GAIA} $\GBP$    & 0.5387   & $1.700\pm0.007$      &                  & $1.002\pm0.007$      &$2.429\pm0.015$ \\                                                                            
{\it GAIA} $\GRP$    & 0.7667   & $1$                  &                  & $0.589\pm0.004$      &$1.429\pm0.015$ \\
Johnson $B$    & 0.4525   & $2.206\pm0.023$      &                  & $1.317\pm0.016$      &$3.151\pm0.027$ \\ 
Johnson $V$    & 0.5525   & $1.675\pm0.010$      &                  & $1            $      &$2.394\pm0.018$ \\ 
SDSS $u$       & 0.3602   & $2.653\pm0.024$      &                  & $1.584\pm0.017$      &$3.791\pm0.028$ \\ 
SDSS $g$       & 0.4784   & $2.018\pm0.012$      &                  & $1.205\pm0.010$      &$2.883\pm0.019$ \\ 
SDSS $r$       & 0.6166   & $1.421\pm0.006$      &                  & $0.848\pm0.006$      &$2.030\pm0.016$ \\ 
SDSS $i$       & 0.7483   & $1.056\pm0.002$      &                  & $0.630\pm0.004$      &$1.509\pm0.015$ \\ 
SDSS $z$       & 0.8915   & $0.767\pm0.004$      &                  & $0.458\pm0.003$      &$1.096\pm0.012$ \\ 
Pan-STARRS $g$ & 0.4957   & $1.934\pm0.010$      &                  & $1.155\pm0.009$      &$2.764\pm0.018$ \\ 
Pan-STARRS $r$ & 0.6211   & $1.413\pm0.005$      &                  & $0.843\pm0.006$      &$2.019\pm0.015$ \\ 
Pan-STARRS $i$ & 0.7522   & $1.052\pm0.001$      &                  & $0.628\pm0.004$      &$1.503\pm0.015$ \\ 
Pan-STARRS $z$ & 0.8671   & $0.815\pm0.002$      &                  & $0.487\pm0.003$      &$1.165\pm0.012$ \\ 
Pan-STARRS $y$ & 0.9707   & $0.662\pm0.004$      &                  & $0.395\pm0.003$      &$0.947\pm0.011$ \\ 
2MASS $J$      & 1.2345   & $0.407\pm0.007$      &                  & $0.243\pm0.004$      &$0.582\pm0.011$ \\ 
2MASS $H$      & 1.6393   & $0.219\pm0.010$      &   $0.222\pm0.012$& $0.131\pm0.006$      &$0.313\pm0.014$ \\ 
2MASS $\Ks$    & 2.1757   & $0.125\pm0.010$      &   $0.130\pm0.006$& $0.078\pm0.004$      &$0.186\pm0.009$ \\ 
{\it WISE} $W1$      & 3.3172   & $0.055\pm0.011$      &   $0.066\pm0.006$& $0.039\pm0.004$      &$0.094\pm0.009$ \\ 
{\it WISE} $W2$      & 4.5501   & $0.029\pm0.011$      &   $0.044\pm0.006$& $0.026\pm0.004$      &$0.063\pm0.009$ \\ 
{\it WISE} $W3$      & 11.7281  & $0.066\pm0.016$      &                  & $0.040\pm0.009$      &$0.095\pm0.021$ \\ 
{\it GAIA} $G$       & 0.6419   & $1.323\pm0.003$      &                  & $0.789\pm0.005$      &$1.890\pm0.015$ \\ 
{\it Spitzer} [3.6]  &          &                      &   $0.062\pm0.005$& $0.037\pm0.003$      &$0.089\pm0.007$ \\   
{\it Spitzer} [4.5]  &          &                      &   $0.044\pm0.005$& $0.026\pm0.003$      &$0.063\pm0.007$ \\   
{\it Spitzer} [5.8]  &          &                      &   $0.031\pm0.005$& $0.019\pm0.003$      &$0.044\pm0.007$ \\
{\it Spitzer} [8.0]  &          &                      &   $0.042\pm0.005$& $0.025\pm0.003$      &$0.060\pm0.007$ \\
\hline
\end{tabular}
\tablenotetext{}{At {\it Spitzer} bands, the determination of the relative extinction $A_\lambda/\Av$ and the extinction coefficient $A_\lambda/E(\GBP-\GRP)$ are based on the relative extinction values from Chen18.}
\end{center}
\end{table}

In the conversion of the CER $E(\lambda - \GRP)/E(\GBP - \GRP)$ into the relative extinction $A_{\lambda}/\ARP$ by Equation~(\ref{equ3}), knowledge of $\ABP/\ARP$ is required. 
Here, we independently calculate $\ABP/\ARP$ by two methods.

Chen18 adopted the color excess--extinction method to determine the extinction along the sight lines toward the Galactic center region by classical Cepheids. 
In that work, they determined the slope of CE versus the absolute extinction plus the relevant distance modulus (DM) diagram and then converted the slope to the relative extinction.
They proved that the extinction law determined by the color excess--extinction method  is consistent with those derived by other methods, including the color--excess method. 
Inspired by that work, we combine the precise parallax and photometric data from the {\it Gaia} catalog to determine $\ABP/\ARP$ by the observed color index $(\GBP - \GRP)$ versus $\GRP$-band apparent magnitude minus DM $\GRP - DM = \ARP + \MRP$  diagram. 
We also refer to this method as the color excess--extinction method. 
If all the RC stars are not affected by extinction, they would appear as a clump with a small scatter in the diagram. 
The RC stars affected by extinction distribute along a linear line.  
Therefore, the relative extinction $\ABP/\ARP$ can be derived from the slope of the linear fit.

As mentioned in Section 2.2, our RC sample contains contaminations, such as SRC stars and RGB stars. 
SRC stars are less luminous than RC stars, which have ignited He in nondegenerate conditions (Girardi 1999).   
In the CMD, SRC stars are located in the bluer and fainter part compared to RC stars. 
In our sample, most SRC stars satisfy $\log g>2.5$ in the APOGEE $\Teff$--$\log g$ contour. 
Therefore, we only select RC stars within $2.35 \le \log g \le 2.5$ to eliminate SRC stars and RGB stars. 
After that, a subsample with 30,431 RC stars is obtained. 
We adopt the distance converted from {\it Gaia} DR2 parallax with corrections by Bailer-Jones et al.\ (2018), who used the Bayesian inference approach to account for the nonlinearity of the transformation and the asymmetry of the resulting probability distribution. 
To reduce uncertainties, we also require the fractional error of the parallax to be less than 0.1 and the parallax to be larger than 0.25. 
After application of these selection criteria, only 5051 RC stars remain. 
Figure~\ref{fig:ABP/ARP} displays the distribution of these RC stars in the $(\GBP - \GRP)$ versus $\GRP - {\rm DM}$ diagram. 
In the range of $(\GBP - \GRP)<1.5$ mag, the uncertainty of DM per extinction is large, and it may affect the result of the linear fit. 
Therefore, we subdivided the sample by the selection cut $(\GBP - \GRP)>1.0+0.001*n$, where $n$ varies from 0 to 500. The linear fit is applied to each subsample.
When $(\GBP - \GRP)>1.341$, the fit achieves the largest coefficient of determination $R^2$,  and we accept it as the best fit. 
It is shown as the black line in Figure~\ref{fig:ABP/ARP}, 
and the slope is $\ARP=(1.429\pm0.015)E(\GBP-\GRP)$, which corresponds to $\ABP/\ARP=1.700\pm0.007$.

For comparison, we also use the color--excess method to determine the extinction $\ABP/\ARP$. 
In section 3.3, we have derived $E(W2-\GRP)/E(\GBP-\GRP)=-1.388$ by fitting the CE--CE diagram. 
The corresponding $E(W2-\GBP)/E(\GBP-\GRP)$ equals $E(W2-\GRP)/E(\GBP-\GRP)-1=-2.388$. 
Then, the relative extinction $\ABP/\ARP$ can be expressed as 
\begin{equation}  \label{equ6}
\ABP/\ARP=\frac{2.388}{1.388+A_{\rm W2}/\ABP}~~.
\end{equation} 
With the assumption of $W2$ band extinction $A_{\rm W2}=0$, we derived the upper limit of $\ABP/\ARP=1.720$. 
$\ABP/\ARP=1.700\pm0.007$ determined from the color--extinction method satisfies this criterion. 
In the literature, the relative extinction $A_{\rm W2}/\AKs$ is $0.34\pm0.10$ (Xue et al.\ 2016; Wang et al.\ 2018; Chen18). 
By combining the $A_{\rm W2}/\AKs$ with the CERs $E(\Ks-\GRP)/E(\GBP-\GRP)$ and $E(W2-\GRP)/E(\GBP-\GRP)$ obtained in Section 3.1, we derive $A_{\rm W2}/\ABP=0.029\pm0.013$. 
Plugging it into Equation~(\ref{equ6}), the derived $\ABP/\ARP=1.686\pm0.016$ from the color--excess method is consistent with that of $1.700\pm0.007$ from color--extinction method.   

Based on the $\ABP/\ARP$ derived here and the CER $k_\lambda$ listed in Table~\ref{tab:cc}, the optical to mid-IR multiband relative extinction $A_{\lambda}/\ARP$ can be estimated by Equation~(\ref{equ3}). 
These results are tabulated in Table~\ref{tab:ext} (Column (3)). 
Chen18 derived the IR extinction law $A_\lambda/\AKs$ in 2MASS, {\it WISE}, and {\it Spitzer} bands using classical Cepheids projected toward the Galactic center region by three methods. 
We converted their relative extinction values into the $A_\lambda/\ARP$ and listed them in Column (4) of Table~\ref{tab:ext}.
Our NIR measurements are in excellent agreement with their results. 
Our mid-IR {\it WISE} results agree with their results at 1.2 $\sigma$. 
Likewise, $A_\lambda/\Av$ (Column (5)) and $A_\lambda/E(\GBP-\GRP)$ (Column (6)) are derived and listed in Table~\ref{tab:ext}. 
Note that the mid-IR {\it WISE} band $A_\lambda/\Av$ results are refined based on more precise values from Chen18. 
Four {\it Spitzer} band extinction values are calculated from Chen18.

\section{Uncertainties}

\subsection{The Residuals of the Color Excess--Color Excess Diagram}

\begin{figure}[h!]
\centering
\vspace{-0.0in}
\includegraphics[angle=0,width=7.0in]{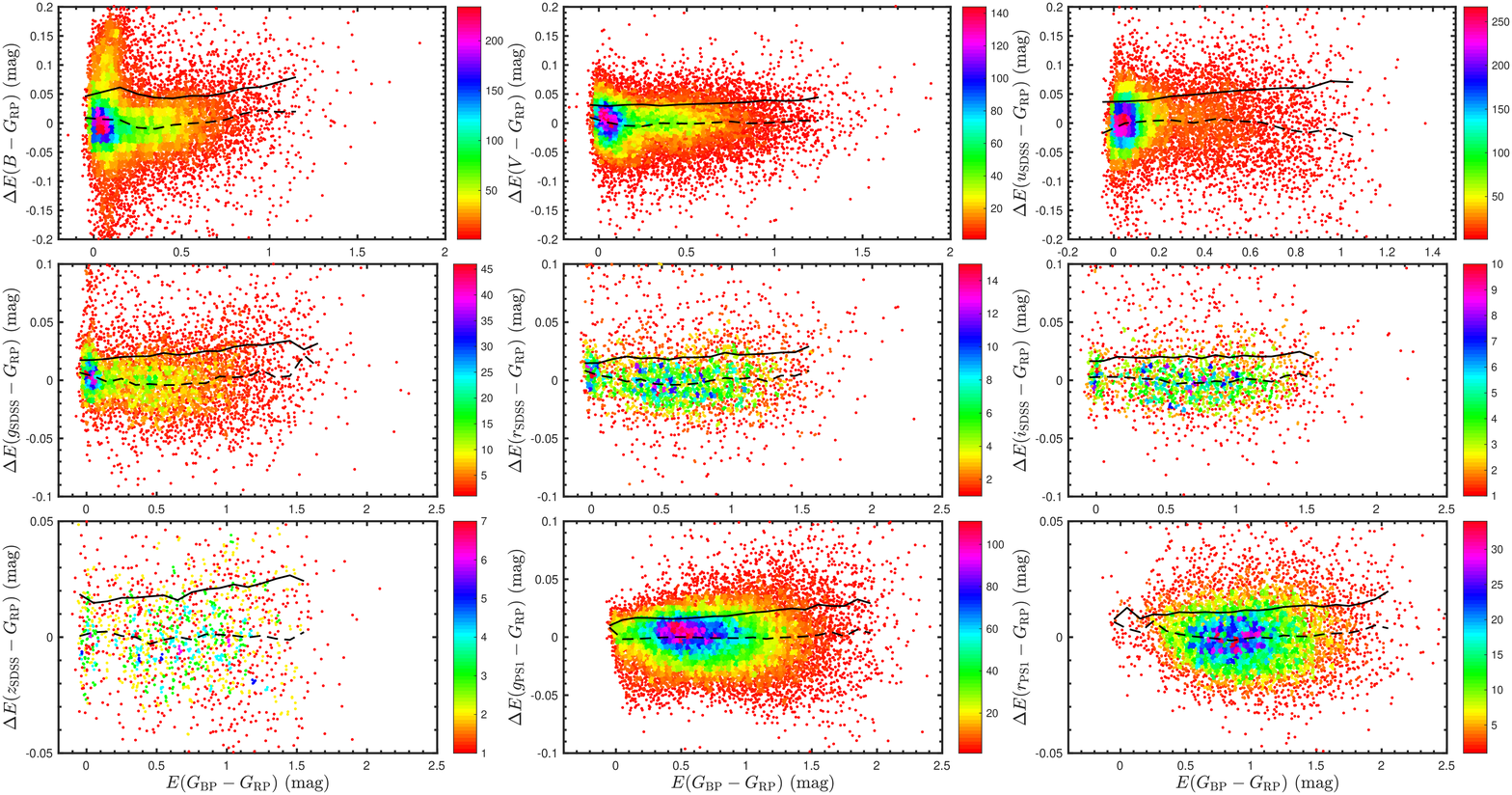}
\vspace{-0.0in}
\caption{\footnotesize
               \label{fig:resid1}
          Distribution of residuals $\Delta E(\lambda - \GRP)$ versus CE $E(\GBP - \GRP)$ 
          diagrams, colored by the number density of RC stars. 
          The residuals are the CE minus the fitted results in Figure~\ref{fig:cc1}. 
          The solid lines and dashed lines are the RMSE 
          and the mean value of residuals ($\overline \Delta$) 
          for stars in bins of $\Delta E(\GBP - \GRP)= 0.1$ mag, respectively. 
               }
\end{figure}
\begin{figure}[h!]
\centering
\vspace{-0.0in}
\includegraphics[angle=0,width=7.0in]{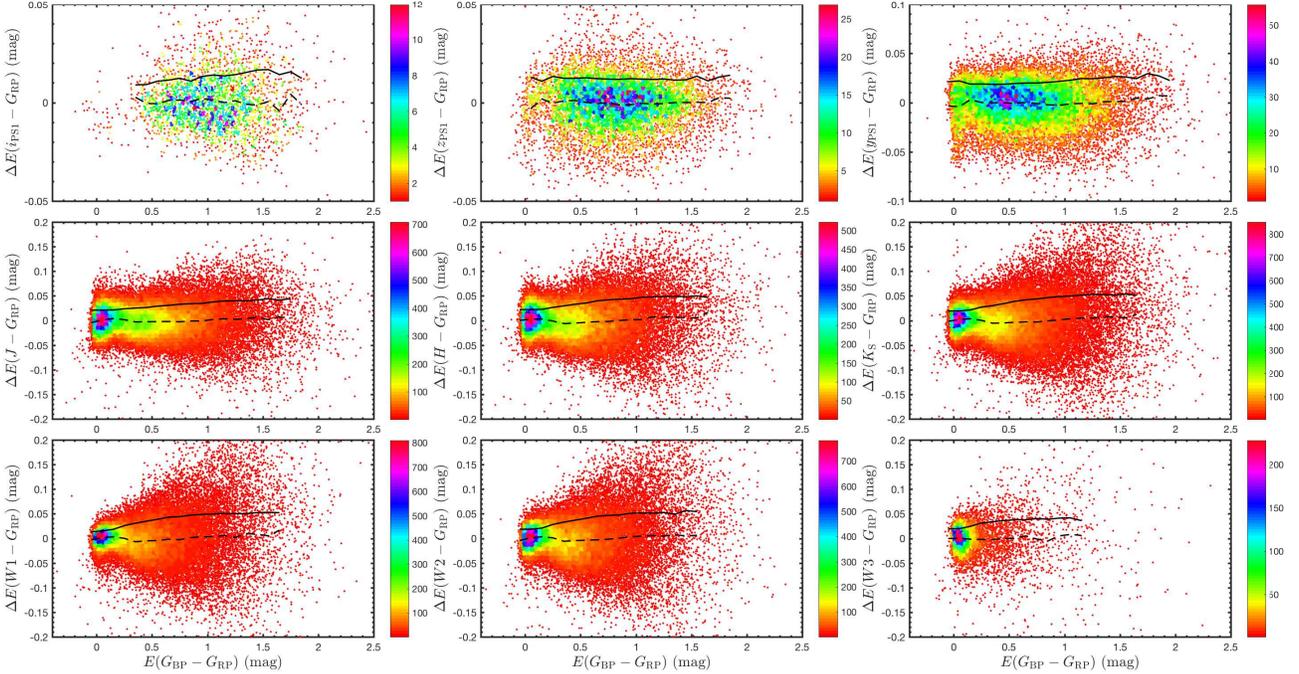}
\vspace{-0.0in}
\caption{\footnotesize
               \label{fig:resid2}
          Same diagram of residuals distribution as in Figure~\ref{fig:resid1}, but the residuals are the CE minus the fitted results in Figure~\ref{fig:cc2}.  
               }
\end{figure}

Although the most RC stars fall right along the fitted lines with only $\lesssim$ 0.001 fitting errors of the slopes, these errors seem to be too small to represent the real errors of CERs. 
In this section, we did a residual analysis to estimate the real errors of CERs. 
To achieve this, we plotted the residual (CE minus fitted functions) $\Delta E(\lambda - \GRP)$ distribution as a function of CE $E(\GBP - \GRP)$ diagrams (Figures~\ref{fig:resid1} and~\ref{fig:resid2}). 
The color shows the number density of RC stars. 
The black solid lines and dashed lines are the RMSE and the mean value of residuals ($\overline \Delta$) for stars in bins of $\Delta E(\GBP - \GRP)= 0.1$ mag, respectively. 
The maxima of the RMSE and the $\overline \Delta$ are listed in the last two columns of Table~\ref{tab:cc}, named as $\rm (RMSE)_{\rm max}$ and $(\overline \Delta)_{\rm max}$.  

At a given band, the photometric uncertainties of the observed color indices and the dispersions of the intrinsic color indices contribute to the residuals $\Delta E(\lambda - \GRP)$. 
As seen in Figures~\ref{fig:resid1} and~\ref{fig:resid2}, 
the average scatter in residuals is small as a whole.  
Among the 21 bands used in this study, residuals in {\it Gaia} bands have the smallest dispersion with $\rm (RMSE)_{\rm max}=0.008$ as shown in Figures~\ref{fig:Gbprp} (b), which is a factor of 2 better than those in PS1 bands.  
The reason is that the data from {\it Gaia} have the lowest photometric uncertainties, and the scatter of intrinsic color of RC stars is small as well. 
In PS1/$grizy$ bands and SDSS/$griz$ bands, the $\rm (RMSE)_{\rm max}$ are around 0.01--0.03. 
In the IR and short-wavelength optical bands, limited by the accuracy of the photometric data, the scatters of the residuals are obviously large. 
For example, the $\rm (RMSE)_{\rm max}$ is around 0.05 mag in 2MASS and {\it WISE} bands and about 0.07 mag in $B$ and $u_{\rm SDSS}$ bands.  
As for $\overline \Delta$, all these bands have a systematic deviation around or much less than 0.02 mag, and it validates the process of the curvature correction.  
We investigate the effects of the RMSE and the $\overline \Delta$ on the CERs and determine the statistical and systematic errors of the CERs in Section 4.2.

\subsection{Simulation}

The $E(\lambda-\GRP)$ versus $E(\GBP-\GRP)$ CE--CE diagrams are simulated to investigate the effects of $x$-axis and $y$-axis error on slopes (CERs).
To obtain the simulated CEs $E(\GBP-\GRP)_{\rm sim}$ and $E(\lambda-\GRP)_{\rm sim}$, we execute a double-Gaussian function to fit the CE $E(\GBP-\GRP)$ in Figures~\ref{fig:cc1} and~\ref{fig:cc2}. 
The first Gaussian distribution represents the low-extinction sources located in the solar neighborhood or high Galactic latitude. 
The second Gaussian distribution represents the high-extinction sources in the disk. 
Hence, the simulation value $E(\GBP-\GRP)_{\rm sim}$ can be expressed as: 
\begin{equation}\label{equ7}
E(\GBP-\GRP)_{\rm sim}=
\frac{1}{\sqrt{2\pi}\sigma_1}\times
\exp\left\{-\left[\frac{(x-\mu_1)^2}{2\sigma_1^2}\right]\right\} 
+
\frac{1}{\sqrt{2\pi}\sigma_2}\times
\exp\left\{-\left[\frac{(x-\mu_2)^2}{2\sigma_2^2}\right]\right\} ~~,
\end{equation}
where $\sigma_1$, $\mu_1$, $\sigma_2$, and $\mu_2$ are the fitting parameters of the double Gaussian function. 
Then, the simulation value $E(\lambda-\GRP)_{\rm sim}$ can be estimated through $E(\lambda-\GRP)_{\rm sim}=k_\lambda E(\GBP-\GRP)_{\rm sim}$, where $k_\lambda$ is the slope of the CE--CE diagram listed in Table~\ref{tab:cc}. 
This process is applied to each band shown in Figures~\ref{fig:cc1} and~\ref{fig:cc2},

First, we test the effects of $x$-axis errors on the slopes. 
The $x$-axis error (i.e. error of the $E(\GBP-\GRP)$) can be inferred from the dispersion of the fit for the CER $E(G-\GRP)/E(\GBP-\GRP)$ (0.005 from Table~\ref{tab:cc}). 
To avoid underestimating the uncertainty, the $x$-axis error is set to be 0.005.
For each band, we generated both the $E(\GBP-\GRP)_{\rm sim}$ and the $E(\lambda-\GRP)_{\rm sim}$ based on the distributions of the observed CEs. 
 After that, by fitting the $E(\GBP-\GRP)_{\rm sim}$ + error versus $E(\lambda-\GRP)_{\rm sim}$ diagrams, we determined the slope $(k_x)_{\rm sim}$, the statistical error of the slope $(\sigma_x)_{\rm sim}$, and the deviation $(\Delta_x)_{\rm sim}$ between the $(k_x)_{\rm sim}$ and the $k_\lambda$. 
Thanks to high-precision {\it Gaia} photometry, the $x$-axis error introduces less than 0.002 deviation to the slopes.

Then, we analyzed the impact of $y$-axis errors on the slopes in a similar way. 
Two $y$-axis errors were considered: 
one is the (RMSE)$_{\rm max}$ representing the local maximum scatter; 
the other one is the nonlinear function that is derived by a polynomial fit to the residuals (dashed lines in Figures~\ref{fig:resid1} and~\ref{fig:resid2}). 
After that, by fitting the $E(\GBP-\GRP)_{\rm sim}$ vs. $E(\lambda-\GRP)_{\rm sim}$ + errors diagrams, we determined the slope $(k_y)_{\rm sim}$, the statistical error of the slope $(\sigma_y)_{\rm sim}$, and the deviation $(\Delta_y)_{\rm sim}$ between the $(k_y)_{\rm sim}$ and the $k_\lambda$. 
By the combination of the $(\sigma_x)_{\rm sim}$ and the $(\sigma_y)_{\rm sim}$, we derived the statistical error, listed as the first error item in the third column of Table~\ref{tab:cc}. 
The sum of the deviations $(\Delta_x)_{\rm sim}$ and $(\Delta_y)_{\rm sim}$ is considered as the systematic error. 
They are listed as the second error item in the third column of Table~\ref{tab:cc}. 
In conclusion, the total uncertainties of the slopes (CERs, Table~\ref{tab:cc}) are mostly less than 0.02.

\begin{figure}[h!]
\centering
\vspace{-0.0in}
\includegraphics[angle=0,width=6.5in]{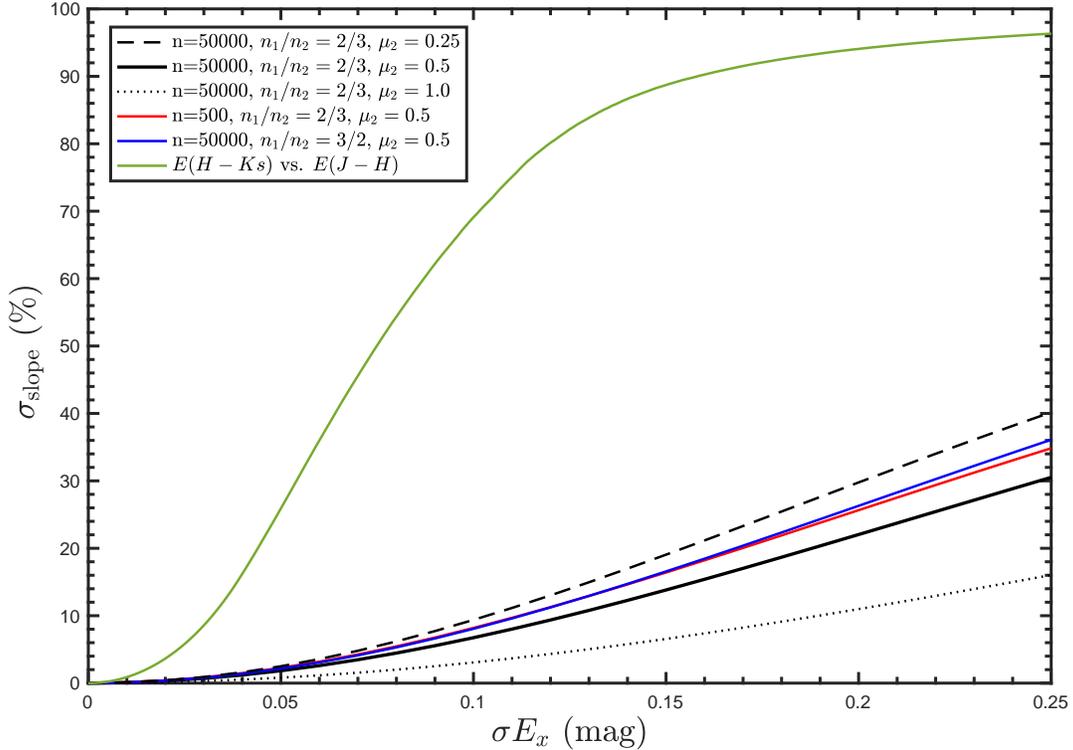}
\vspace{-0.0in}
\caption{\footnotesize
               \label{fig:sim}
          The percentage deviation of the slope $\sigma_{\rm slope}$ varies 
          as the $x$-axis CE error $\sigma E_x$.  
          Five cases are considered with different parameters $n$, $n_1/n_2$, and $\mu_2$, 
          where $n$ is the total number of objects, 
          $n_1/n_2$ is the ratio of sources with low and high extinction, 
          and $\mu_2$ is the average reddening amount of high-extinction sources. 
          The green line shows the similar simulation for 
          about 60,000 2MASS-APOGEE RCs in $E(H-\Ks)$ vs. $E(J-H)$.  
               }
\end{figure}

In addition, we analyzed the effects of various $x$-axis CE errors $\sigma E_x$ on the slopes. 
This process can explain why we adopt Gaia bands as the $x$-axis.
We model the CE--CE diagram of 2MASS $\Ks$ band as an example.    
The corresponding values of the parameters $\sigma_1$, $\mu_1$, $\sigma_2$, and $\mu_2$ are determined. 
Then, we considered five cases with various parameters ($n$, $n_1/n_2$, and $\mu_2$) to represent different initial conditions. 
More specially, the total number of objects $n$ is 500 or 50,000, 
the ratio of low- to high-extinction sources $n_1/n_2$ is 2/3 or 3/2, 
and the average reddening amount of high-extinction sources $\mu_2$ is 0.25, 0.5, or 1.0 mag. 
Large $\mu_2$ denotes high extinction.   
Lastly, a group of $x$-axis CE errors $\sigma E_x$ ranging from 0.0 to 0.25 was added to CE $E(\GBP-\GRP)_{\rm sim}$. 
By fitting the $E(\GBP-\GRP)_{\rm sim} + \sigma E_x$ vs. $E(\Ks-\GRP)_{\rm sim}$ diagram, we derived the slope $(k_\Ks)_{\rm sim}$. 
In comparison to the slope $k_\Ks$ of $E(\Ks-\GRP)/E(\GBP-\GRP)$ determined in Section 3.2, we estimated the percentage deviation of the slope $\sigma_{\rm slope}$ between these two slopes by $\mid ((k_\Ks)_{\rm sim}-k_\Ks)/k_\Ks\mid$.

Figure~\ref{fig:sim} shows that the error of the slope $\sigma_{\rm slope}$ varies as the CE error $\sigma E_x$. 
Generally, the $\sigma_{\rm slope}$ increases with the increase of $\sigma E_x$. 
For $\sigma E_x < 0.02$ mag, the slowly increasing errors of the slopes are small and almost similar in different cases, 
while for $\sigma E_x > 0.06$ mag, the errors of the slopes increase dramatically.
As shown in the black dashed, solid, and dotted lines, the increase of $\mu_2$ could effectively reduce the error of the slope.  
In addition, the $\sigma_{\rm slope}$ of the red line with $n=500$ is larger than that of the solid black line with $n=$50,000. 
This means that increasing the total number of objects can reduce the error of the slope. 
Furthermore, we analyzed the influence caused by the ratio of low- to high-extinction sources $n_1/n_2$. 
The $\sigma_{\rm slope}$ of the blue line with $n_1/n_2$ = 3/2 is slightly higher than that of the solid black line with $n1/n2$ = 2/3 from beginning to end.   
This means that the higher the proportion of high-extinction sources, the smaller the error of the slope. 
In conclusion, the error of the slope increases as the $x$-axis CE error increases. 
For the purpose of improving the precision of CER, the bands with the best photometric quality should be set as the basis $x$-axis bands. 
When the $x$-axis error $\sigma E_x$ is notable, three parameters, average reddening amount of high-extinction sources $\mu_2$, total number of objects $n$, and ratio of low- to high-extinction sources $n_1/n_2$, will also exacerbate uncertainties of the extinction law.

Figure~\ref{fig:sim} can also be used to explain the superiority of using {\it Gaia} bands rather than 2MASS bands as the basis of CER analysis. 
We redo a similar analysis of $E(H-\Ks)$ vs. $E(J-H)$ based on about 60,000 2MASS-APOGEE RCs, and the result is shown as the green line in Figure~\ref{fig:sim}. The error of the slope is several times larger than that caused by adopting {\it Gaia} as basis bands. If we adopt the $\sigma E_x$ = 0.03 mag for the 2MASS CE, the corresponding 1$\sigma$ error of the CER is around 8.6\%. 
Besides, when the $x$-axis error $\sigma E_x$ is large, the other parameters, including the average reddening value of high-extinction sources and the sample number, will also have a great impact on the slope (CER). 
In addition, as mentioned in Section 3.3 and Stead \& Hoare (2009), for high-extinction regions, such as the Galactic center and molecular clouds, the curvature of CER becomes obvious. 
The curvature of CERs also influences the measurement of the slope. 
In conclusion, all of these are the reasons why various NIR CERs were reported in previous works (Section 1.2). 
Compared to the error in the $E(J-H)$, the error in the $E(\GBP-\GRP)$ is much smaller ($< 0.005$ mag) (Figure~\ref{fig:resid1}). 
No matter what the situation is, the effect on the error of the slope is less than 1\% and could be ignored. 
Therefore, we recommend adopting high photometric precision bands (here, $\GBP$ and $\GRP$) as basis bands in CER analysis to reduce the error of the slope caused by the fitting method.

\section{Discussion}

\subsection{Extinction of Individual RC Sight Lines}

The RC star is a standard candle, as its luminosity has relatively week dependency on the stellar composition, color, and age in the solar neighborhood (Paczy\'nski \& Stanek 1998; Alves 2000; Groenewegen 2008; Girardi 2016). 
The wavelength-dependent extinction to each RC star sight line can be estimated by the formula $A_\lambda = m_\lambda - M_\lambda - 5\log d + 5$ with the known values of apparent magnitude $m_\lambda$, absolute magnitude $M_\lambda$, and distance $d$. 
For our RC sample, with the multiband apparent magnitudes $m_\lambda$ from the photometric catalogs and the distance information $d$ from the {\it Gaia} DR2 catalog, we only need the absolute magnitude $M_\lambda$ to estimate the extinction $A_\lambda$.
Compared to optical bands, the absolute magnitude of RC stars has a reduced systematic dependence on metallicity and effective temperature in IR bands. 
Among the RC absolute magnitudes given in the literature, the $\Ks$-band absolute magnitude $M_{\Ks}$ has the most consistent value (Ruiz-Dern et al.\ 2018 and reference therein).

As our goal is to determine the accurate extinction laws toward RC sight lines, 
we evaluate the feasibility of this single-star sight-line method through error analysis. 
To achieve this, the fractional error of the extinction $(A_\lambda)_{\rm err}/A_\lambda$ is estimated at the given $\lambda$ band.  
The extinction error $(A_\lambda)_{\rm err}$ comes from the errors of photometry $(m_\lambda)_{\rm err}$, the absolute magnitude $(M_\lambda)_{\rm err}$, and the parallax $d_{\rm err}$. 
Two photometric bands are taken as examples, the optical SDSS $g$ band, which has larger extinction than IR bands, and the IR 2MASS $\Ks$ band, which has the most consistent $M_{\Ks}$.

The typical photometric error is $\sim0.03$ mag at $g$ and $\Ks$ bands. 
We select RCs by the criteria of distance less than 4 kpc and the fractional error of the parallax less than 0.1. The average error of the distance to the RC stars is $\sim0.18$ mag. 
The absolute magnitudes of the RC stars are $M_g=1.229\pm0.172\magni$ (Chen et al.\ 2017) and $M_{\Ks}=-1.61\pm0.03\magni$ (Alves 2000; Ruiz-Dern et al.\ 2018). 
Therefore, the typical extinction errors $(A_\lambda)_{\rm err}$ at $g$ and $\Ks$ bands are 0.19 and 0.25$\magni$. 
The average extinction values of individual RC sight lines at $g$ and $\Ks$ bands are $A_g=2.8\magni$ and $\AKs=0.2\magni$, respectively. 
The extinction uncertainty of a single RC sight line is $\sim9$\% at $g$ band and reaches $\sim100$\% at $\Ks$ band. 
Compared to the accurate extinction law determined by the color--excess method (Table~\ref{tab:cc}), the extinction uncertainties of this method are significant. 
Therefore, the current single-star sight-line analysis is inferior to the statistical color--excess method and we only treat the RC targets as a whole to investigate the extinction law.

\subsection{The Comparison of Color Excess Ratios}

\begin{table}[h!]
\begin{center}
\caption{\label{tab:comp} Color Excess Ratios Compared to Previous Works}
\vspace{0.1in}
\small
\begin{tabular}{lcccc}
\hline \hline 
Color Excess Ratios        & This Work        &  Work (a)       &  Work (b)       &  Work (c) \\ 
 \hline
SDSS  $E(u-g)/E(g-r)$      & $1.064\pm0.013$  & $1.010\pm0.020$ & $1.091\pm0.019$             & ...\\                                                                       
SDSS  $E(g-r)/E(r-i)$      & $1.637\pm0.011$  & $1.695\pm0.057$ & $1.650\pm0.044$ & ...\\
SDSS  $E(r-i)/E(i-z)$      & $1.261\pm0.010$  & $1.299\pm0.034$ & $1.395\pm0.034$ & ...\\
SDSS  $E(i-z)/E(z-J)$      & $0.804\pm0.006$  & ...             & $0.768\pm0.020$ & ...\\
PS1   $E(g-r)/E(r-i)$      & $1.445\pm0.004$  & ...             & ...             & $1.395\pm0.013$ \\
PS1   $E(r-i)/E(i-z)$      & $1.524\pm0.007$  & ...             & ...             & $1.531\pm0.012$ \\
PS1   $E(i-z)/E(z-J)$      & $0.580\pm0.002$  & ...             & ...             & $0.558\pm0.008$ \\
2MASS $E(J-H)/E(H-\Ks)$    & $2.006\pm0.046$  & ...             & $1.625\pm0.063$ & $1.943\pm0.019$ \\
2MASS $E(H-\Ks)/E(\Ks-W1)$ & $1.344\pm0.036$  & ...             & $1.333\pm0.094$ & $1.348\pm0.040$ \\
{\it WISE}  $E(\Ks-W1)/E(W1-W2)$ & $2.629\pm0.152$  & ...             & $4.615\pm1.428$ & $2.627\pm0.187$ \\
\hline
\end{tabular}
\normalsize
\tablenotetext{}{Previous works include (a) Schlafly et al.\ (2011), 
(b) Yuan et al.\ (2013), and (c) Schlafly et al.\ (2016). }
\end{center}
\end{table}

The CERs have been measured by a number of works in a variety of photometric bands. 
Hence, we compare our measurements with some measurements reported in the literature. 
In order to compare with publications in different combinations of bands $E(\lambda_1 - \lambda_2)/E(\lambda_2 - \lambda_3)$, 
we convert our CERs (Table~\ref{tab:cc}) into the CERs in the corresponding bands (Table~\ref{tab:comp}). 
For example, we use three CERs $E(u-\GRP)/E(\GBP-\GRP)$, $E(g-\GRP)/E(\GBP-\GRP)$, and $E(r-\GRP)/E(\GBP-\GRP)$ to calculate the CER $E(u-g)/E(g-r)$. 
The errors of these three CERs are propagated to the $E(u-g)/E(g-r)$ as well. 
The CERs derived in this work and reported in previous works are tabulated in Table~\ref{tab:comp}.

As a whole, the agreement between our measurements and those of the previous publications is excellent. 
It is worth noting that the precision of our CERs is better than the previous results. 
Our SDSS band results agree well with Schlafly \& Finkbeiner (2011, work (a)). 
We compare seven CERs through optical to IR bands to the results of Yuan et al.\ (2013, work (b)). 
Four of them are in agreement with each other within 2$\sigma$. 
The agreement between ours and those of Yuan et al.\ (2013) in CERs $E(r-i)/E(i-z)$ and $E(J-H)/E(H-\Ks)$ is out of 3$\sigma$. 
But these two agree with work (a) and Schlafly et al.\ (2016, work (c)) in 2$\sigma$. 
The {\it WISE} result $E(\Ks-W1)/E(W1-W2)$ of Yuan et al.\ (2013) is nearly twice that of ours. 
Overall, our CERs are closely consistent with those derived by Schlafly et al.\ (2016, work (c)), 
who also reported the difference between their results and Yuan et al.\ (2013) in the CERs $E(r-i)/E(i-z)$, $E(J-H)/E(H-\Ks)$, and $E(\Ks-W1)/E(W1-W2)$. 
Schlafly et al.\ (2016) pointed out the measurement in Yuan et al.\ (2013) was uncertain because only the low-reddening objects are available in their sample. 

Our measurements rely on a large sample of RC stars with precise photometry and homogeneous stellar parameters. 
Even though each CER listed in Table~\ref{tab:comp} contains propagated errors of three CERs, our measurements are in great agreement with previous works. 
More importantly, the precision of our CERs has significantly improved.

\subsection{The Comparison with Extinction Curves}

\begin{figure}[h!]
\centering
\vspace{-0.0in}
\includegraphics[angle=0,width=6.5in]{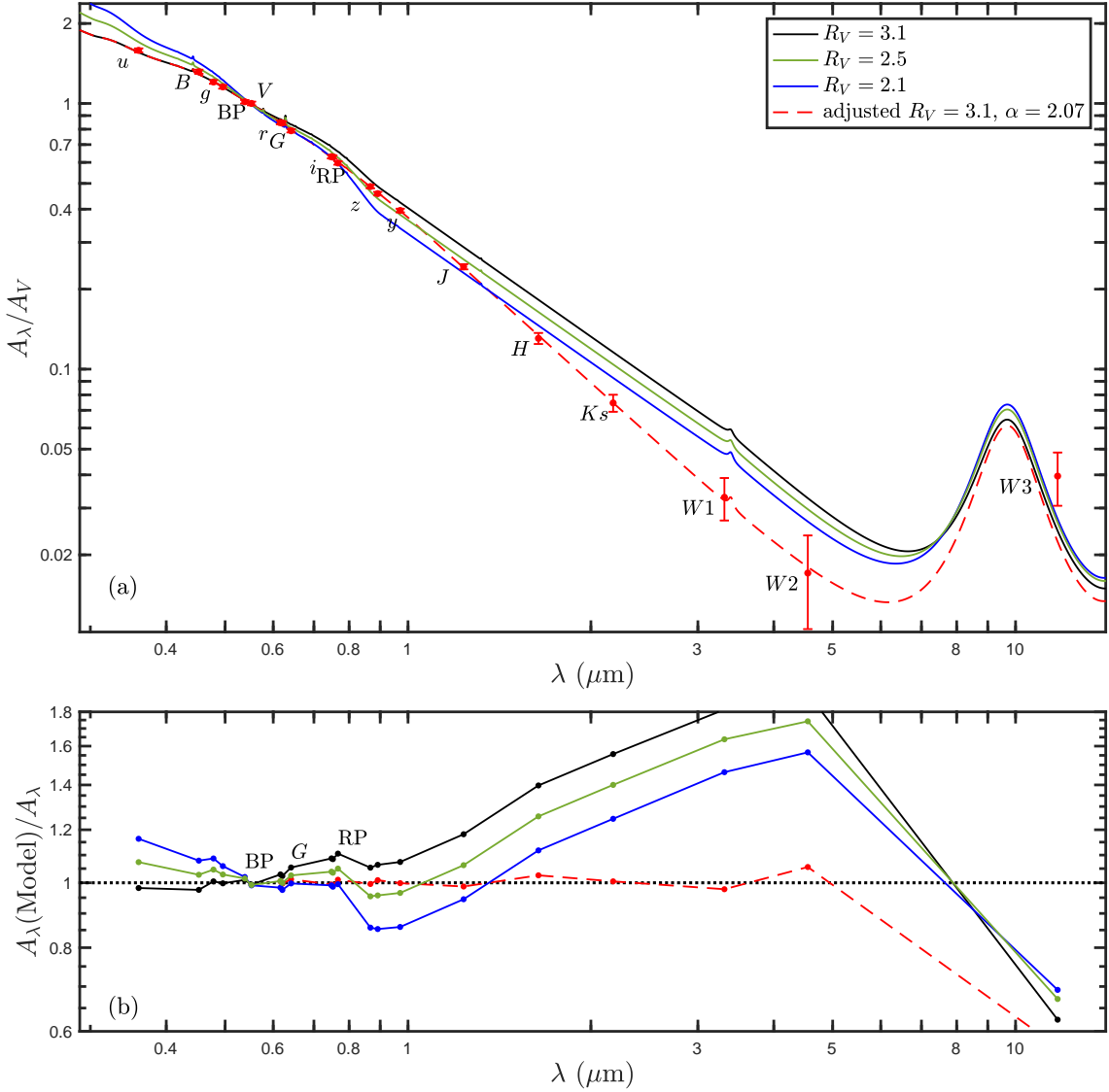}
\vspace{-0.0in}
\caption{\footnotesize
               \label{fig:AAv}
          (a): Optical to mid-IR multiband extinction 
          $A_{\lambda}$ relative to $\Av$ (red circles with error bars). 
          For comparison, the CCM $\Rv$ = 3.1 (black line), 2.5 (green line), 
          2.1 (blue line) model extinction curves are also shown. 
          The observed extinction law can be fitted by the adjusted CCM $\Rv=3.1$ curve 
          (red dashed line).  
          The NIR extinction in the range of $1.0\mum\le\lambda<3.33\mum$ 
          can be well represented by a power-law formulation with the index $\alpha=2.07\pm0.03$. 
          (b): Comparison of model extinction curves to the observed extinction law in ratio.
               }
\end{figure}

To compare with model extinction curves characterized by $\Rv$, we convert our relative extinction results $A_{\lambda}/\ARP$ to $A_{\lambda}/\Av$ and calculate $\Rv$ value as well. 
The extinction values $A_{\lambda}/\ARP$ and $A_{\lambda}/\Av$ are list in the third and fifth columns of Table~\ref{tab:ext}. 
According to the definition of $\Rv$, the total-to-selective extinction ratio, we derive $\Rv=\AV/(\AB-\AV)=1/(\AB/\AV-1)=3.16\pm0.15$. 
As this work has no bias toward any specific environment and covers all the fields surveyed by APOGEE, our measurements represent the average extinction. 
This is in agreement with the average value of the Galactic diffuse ISM $\RV=3.1$.
Fitzpatrick \& Massa (2009) suggested that the CER $E(K-V)/E(B-V)$ can be used to estimate $\RV$ with 0.12 uncertainty. 
Therefore, we estimated the $\Rv$ value by this method as well. 
The determined $\Rv=3.19$ is also consistent with the $\Rv$ determined from the total-to-selective extinction ratio. 
Schlafly et al.\ (2016) used the CER $E(g_{\rm PS1}-W2)/E(g_{\rm PS1}-r_{\rm PS1})$ to obtain the $R'_{\rm V}$ as a proxy of $\RV$. 
Based on the definition, we derive $R'_{\rm V}=3.21$ from our measurements, which is consistent with the average value 3.32 from Schlafly et al.\ (2016). 
Our $\RV$ is determined in static wavelengths (reddening/extinction after the curvature correction), so a slight difference exists when comparing with the previous $\RV$ values.

As the calculation of $\Rv$ only depends on the extinction between two optical bands, or the CER of three bands, this value cannot perfectly reflect the wavelength-dependent extinction law. 
Therefore, we compare our optical to mid-IR multiband extinction with different CCM model extinction curves in Figure~\ref{fig:AAv}. 
For each band, we uniformly calculate the static effective wavelength $\lambda_{\rm eff, 0}$ through 
\begin{equation} \label{equ8}
\lambda_{\rm eff, 0}=\frac{\int \lambda F_\lambda(\lambda)S(\lambda)d\lambda} 
{\int F_\lambda(\lambda)S(\lambda)d\lambda}~~.  
\end{equation}
As mentioned in Section 3.3, the synthetic stellar spectra $F_\lambda(\lambda)$ are based on an RC star with $T_{\rm eff}=4800$ K, $\log g=2.5$, and [Fe/H] $= -0.1$ (Lejeune et al.\ 1997), and $S(\lambda)$ is the filter transmission curve. 
The calculated effective wavelengths $\lambda_{\rm eff, 0}$ are tabulated in the second column of Table~\ref{tab:ext}. 
In Figure~\ref{fig:AAv} (a), our extinction results are plotted as red circles with error bars. 
The CCM $\Rv$ = 3.1 (black line), 2.5 (green line), and 2.1 (blue line) model extinction curves are also shown. 
Figure~\ref{fig:AAv} (b) is the ratio of model extinction values from CCM $\Rv=3.1$ (black), 2.5 (green), and 2.1 (blue) to our observed extinction values $A_\lambda$ (Model)/$A_\lambda$ at the given band.

As shown in Figure~\ref{fig:AAv}, our optical extinction law conforms with the $\Rv=3.1$ extinction curve ($< 2\%$ deviation) in the wavelength range of 300--550 nm, 
while at longer wavelengths the new determined extinction law is significantly steeper than the $\Rv=3.1$ extinction curve. 
More specifically, it agrees with the $\Rv=2.1$ extinction curve in the wavelength range of 600--770 nm 
and agrees with $\Rv=2.5$ in the range of 770--1000 nm. 
This feature is consistent with the $\RV \approx 2.5$ reported by Nataf et al.\ (2013), who investigated the extinction law toward the Galactic bulge based on $V$ and $I$ bands. 
In NIR bands ($0.9\mum < \lambda < 3\mum$), none of these models can perfectly explain the quite steep trend, so a larger power-law index $\alpha$ is needed. 
To better describe the observed extinction law, we made some adjustments on the CCM $\Rv=3.1$ extinction curve to derive the new equations given below.
 
Optical: $0.3\mum <\lambda< 1.0\mum$ and $Y=1/\lambda(\mum)-1.82$,
\begin{eqnarray} \label{equ9}
A_{\lambda}/\AV  & = & 1.0+0.7499Y-0.1086Y^2-0.08909Y^3+0.02905Y^4 \nonumber\\
                             & &  +0.01069Y^5+0.001707Y^6-0.001002Y^7 ~~; 
\end{eqnarray}

NIR: $1.0\mum \leq \lambda< 3.33\mum$,
\begin{equation} \label{equ10}
A_{\lambda}/\Av = (0.3722\pm0.0026) \lambda^{-2.070\pm0.030}~~. 
\end{equation}
These equations obey the form of the CCM model, while the coefficients are reanalyzed, through best fitting to our extinction in 19 optical and NIR bands, including $u, B$, $g_{\rm SDSS}$, $g_{\rm PS1}$, $\GBP$, $V$, $r_{\rm SDSS}$, $r_{\rm PS1}$, $G$, $i_{\rm SDSS}, i_{\rm PS1}$, $\GRP$, $z_{\rm PS1}$, $z_{\rm SDSS}$, $y_{\rm PS1}$, $J, H, \Ks$, and $W1$ bands. 
The adjusted extinction curve, shown as the red dashed line in Figure~\ref{fig:AAv}, is consistent with the observed extinction law by better than 2.5\%. 
Note that our new extinction law (Equations~(\ref{equ9}) and~(\ref{equ10})) is the continuous extinction curve between 0.3 and $3.33\mum$. 
The IR absorption features presenting at $\gtrsim 3\mum$ (e.g., Draine 2003; Fritz et al.\ 2011, and references therein), such as ice, hydrocarbons features, are not contained in the extinction curve.

The Galactic center is an ideal region to investigate the IR extinction owing to the well-measured galactocentric distance (Bland-Hawthorn \& Gerhard 2016; de Grijs \& Bono 2016) and relatively high extinction. 
Along the Galactic center sight lines, steep NIR extinction laws also have been reported in the literature. 
Sch\"odel et al.\ (2010) derived $\alpha=2.21$ by measuring extinction between $H$ and $K$ bands for RC stars. 
Fritz et al.\ (2011) found $\alpha=2.11\pm0.06$ via measurement of hydrogen emission lines.  
Most recently, Chen18 reported $\alpha=2.05\pm0.07$ traced by classical Cepheids.  
An even steeper Galactic center extinction law has been reported by Nogueras-Lara et al.\ (2018) with $\alpha = 2.31\pm0.03$. 
Generally, our index value $\alpha=2.07\pm0.03$ is consistent with those of Fritz et al.\ (2011) and Chen18. 
At the same time, 
due to much better photometry and parallax of $\it Gaia$, a purer sample of RC stars from APOGEE, and a robust determination method, 
we recommend a steep average NIR extinction with $\alpha=2.07$ instead of the CCM extinction with $\alpha=1.61$ in future work.

\subsection{The {\it Gaia} Extinction Coefficient}

\begin{figure}[h!]
\centering
\vspace{-0.0in}
\subfigure[original]{
\includegraphics[angle=0,width=3.1in]{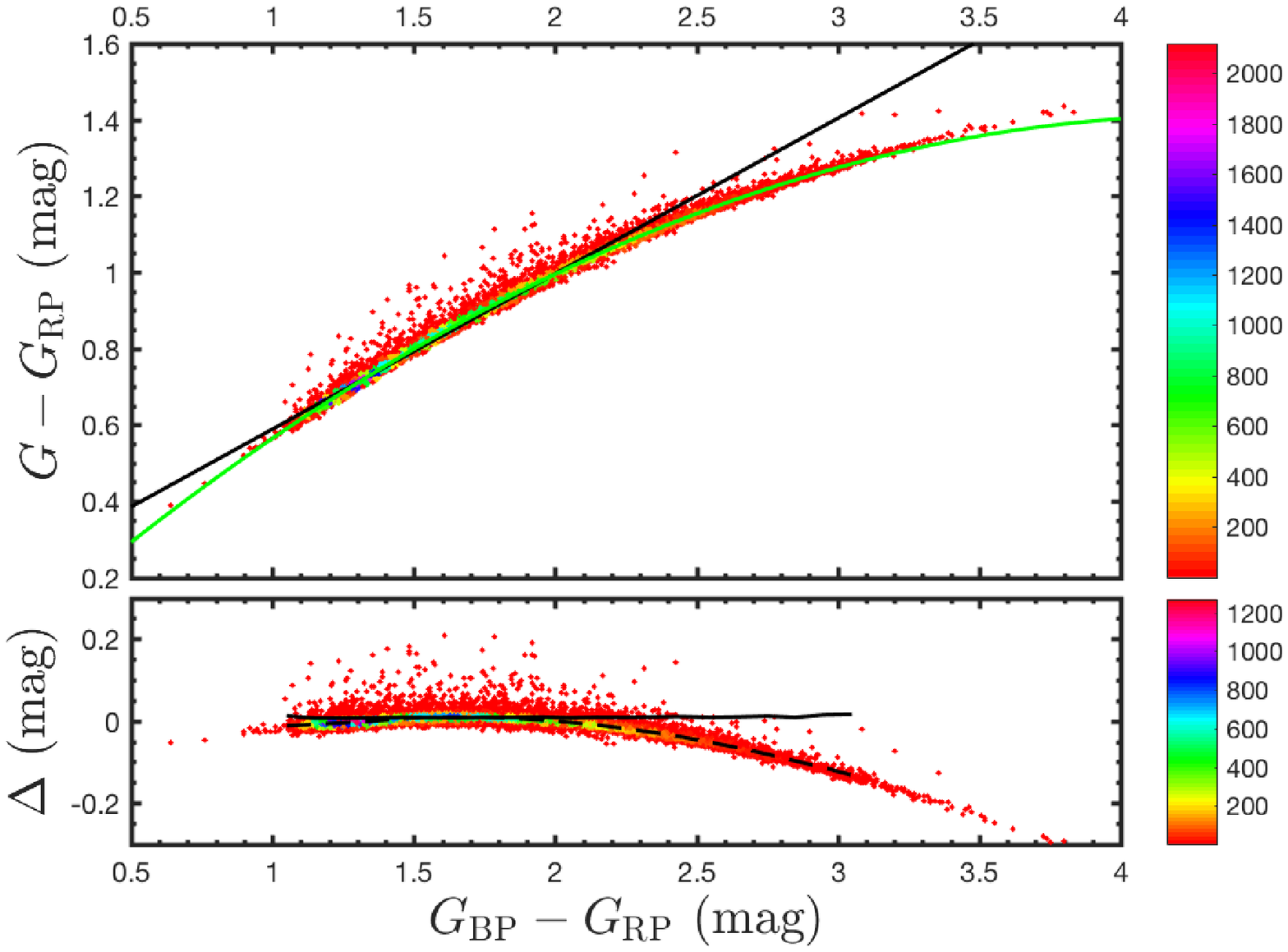}
}
\subfigure[corrected]{
\includegraphics[angle=0,width=3.1in]{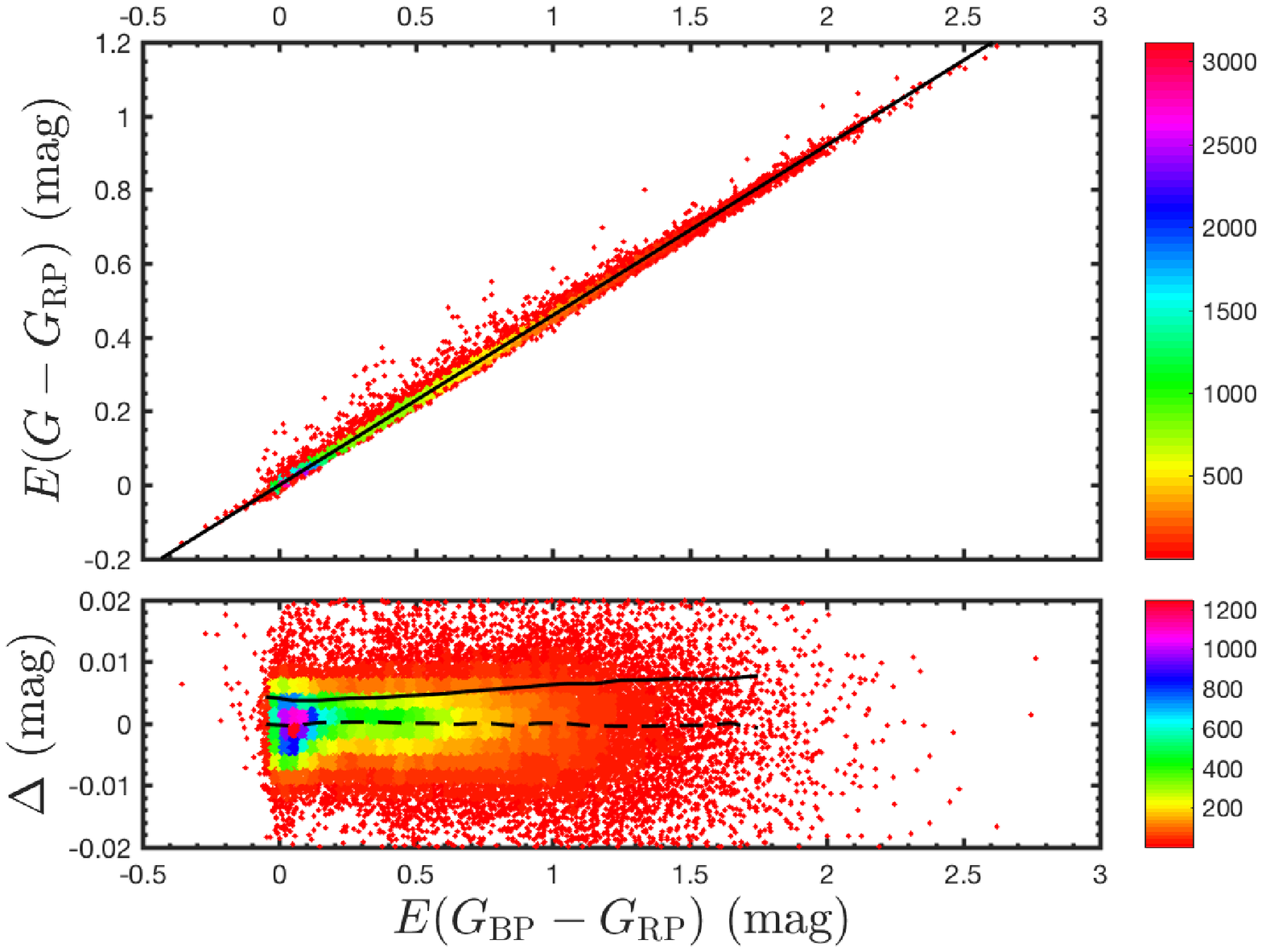}
}
\vspace{-0.0in}
\caption{\footnotesize
               \label{fig:Gbprp}
          The upper panel of (a) is the original 
          color ($\GBP - \GRP$) vs. color ($G - \GRP$) diagram for RC stars. 
          The black and the green lines are the linear fit  
          and second-order polynomial fitting curve, respectively.  
          The upper panel of (b) is the CE $E(\GBP - \GRP)$ vs. CE $E(G - \GRP)$ diagram 
          for RC stars after the curvature correction. 
          The black line is the best linear fit. 
          The lower panels of (a) and (b) display the distributions of residuals. 
          The solid line and the dashed line are the RMSE 
          and the mean value of the residuals, respectively.  
          The color shows the number density of RC stars.  
               }
\end{figure}

Figure~\ref{fig:Gbprp} exhibits the determination of the {\it Gaia} extinction coefficient. 
As shown in Figure~\ref{fig:Gbprp} (a), the distribution of RC stars in the color--color diagram exhibits good linearity and can be fitted by the black straight line up to about $\GBP-\GRP = 2.0$ mag.  
However, at color $\GBP - \GRP \gtsim 2.0$, the distribution begins to curve. 
It displays great amounts of curvature for heavily reddened objects. 
The curve feature is even clearer in the residual distribution diagram (lower panel of Figure~\ref{fig:Gbprp} (a)). 
As discussed in Section 3.3, the curvature of CER is due to the assumption of the static wavelength. 
Because of the broad bandwidth in the {\it Gaia} bands, the curvature is particularly evident. 
To obtain the real extinction at the {\it Gaia} bands, we correct the observed curvature by using our model extinction tracks (Figure~\ref{fig:extcorr}). 
Figure~\ref{fig:Gbprp} (b) displays the CE--CE diagram after the curvature correction. 
The distribution of RC stars exhibits extremely good linearity and is best fitted by the black line with $E(G-\GRP)/E(\GBP-\GRP)=0.461\pm0.000$. 
Combined with the relative extinction $\ABP/\ARP=1.700$ derived in Section 3.4, the {\it Gaia} band extinction coefficients are obtained:  
$\ABP=(2.429\pm0.015)E(\GBP-\GRP)$, $\AG=(1.890\pm0.015)E(\GBP-\GRP)$, 
and $\ARP=(1.429\pm0.015)E(\GBP-\GRP)$, listed in Table~\ref{tab:ext} as well.

We compare {\it Gaia} band extinction values with extinction in some other optical bands in Figure~\ref{fig:AAv}. 
Our optical continuous extinction curve smoothly varies as wavelength.  
The extinctions in $\GBP, G$, and $\GRP$ bands are close to those in $V, r$, and $i$ bands, respectively, as their static effective wavelengths are close to each other (Table~\ref{tab:ext}).  
This agreement proves the reliability of our {\it Gaia} extinction results.

It is worth noting that our extinction law in Table~\ref{tab:ext} and Figure~\ref{fig:AAv} represents the static extinction law. 
Because of the nonignorable bandwidth of the filter, the existence of curvature in reddening/extinction is unavoidable.  
This curvature depends on the spectral type, the filter system, and the amount of extinction. 
Since this curvature is exacerbated with the increase of CE or extinction (Figure~\ref{fig:extcorr}), 
to avoid systematic error, extinction laws are only suitable to estimate and correct extinctions for low-extinction objects. 
For objects with moderate or heavy extinction, the extinction law needs a small correction, 
that is, $R_{\lambda, c}=R_\lambda*A_\lambda/A_{\lambda, 0}$,  
where $\lambda$ denotes the band of interest and  
$R_{\lambda, c}$ and $R_\lambda$ are the corrected and the static extinction law, respectively. 
$A_\lambda$ and $A_{\lambda, 0}$ are the evolving and the static band extinction, respectively, estimated by the combination of the static extinction law, the synthetic stellar spectra, and the filter transmission curve (see details in Section 3.3). 
Similarly, for the correction of the relative extinction, the formula is 
$A_{\lambda_1, c}/A_{\lambda_2, c}=A_{\lambda_1}/A_{\lambda_2}$ (different from  $A_{\lambda_1, 0}/A_{\lambda_2, 0}$), 
where $\lambda_1$ and $\lambda_2$ are two bands of interest.

\subsection{The Predicted Extinction in {\it HST} WFC3 and {\it JWST} NIRCAM Bandpasses}

\begin{table}[h!]
\begin{center}
\caption{\label{tab:JWST} The Predicted {\it HST} and {\it JWST} Extinction Values}
\vspace{0.1in}
\begin{tabular}{cccc|cccc}
\hline \hline    
   Band & $\lambda_{\rm eff}$ ($\Angstrom$) & $A_\lambda/A_V$ & $A_\lambda/A_\Ks$ 
& Band & $\lambda_{\rm eff}$ ($\Angstrom$) & $A_\lambda/A_V$ & $A_\lambda/A_\Ks$\\
\hline
WFC3 F098M & 9849.7 & 0.3835 &	4.9421  &  JWST F140M& 14040.2  &   0.1849 &	2.3832   \\
WFC3 F127M &12736.1 & 0.2252 &	2.9013  &  JWST F162M& 16249.1  &   0.1361 &	1.7535   \\
WFC3 F139M &13833.9 & 0.1909 &	2.4593  &  JWST F182M& 18393.2  &   0.1058 &	1.3637   \\
WFC3 F153M &15316.1 & 0.1543 &	1.9887  &  JWST F210M& 20915.2  &   0.0811 &	1.0445   \\
WFC3 F105W &10438.9 & 0.3417 &	4.4036  &  JWST F250M& 25008.5  &   0.0562 &	0.7236   \\
WFC3 F110W &11169.7 & 0.2966 &	3.8218  &  JWST F300M& 29817.8  &   0.0394 &	0.5072   \\
WFC3 F125W &12335.5 & 0.2417 &	3.1143  &  JWST F335M& 33537.6  &   0.0321 &	0.4136   \\
WFC3 F140W &13692.3 & 0.1939 &	2.4984  &  JWST F360M& 36151.2  &   0.0270 &	0.3481   \\
WFC3 F160W &15258.3 & 0.1556 &	2.0044  &  JWST F410M& 40720.0  &   0.0217 &	0.2801   \\            
JWST F070W & 7040.1 & 0.6919 &	8.9158  &  JWST F430M& 42775.2  &   0.0199 &	0.2570   \\
JWST F090W & 9004.5 & 0.4523 &	5.8277  &  JWST F460M& 46267.6  &   0.0175 &	0.2255   \\
JWST F115W &11503.9 & 0.2785 &	3.5884  &  JWST F480M& 48122.9  &   0.0165 &	0.2129   \\
JWST F150W &14940.9 & 0.1618 &	2.0848  &  JWST F164N& 16450    &   0.1329 &	1.7127   \\
JWST F200W &19694.7 & 0.0919 &	1.1841  &  JWST F187N& 18740    &   0.1018 &	1.3113   \\
JWST F277W &27288.9 & 0.0470 &	0.6054  &  JWST F212N& 21210    &   0.0786 &	1.0123   \\
JWST F356W &35289.7 & 0.0283 &	0.3648  &  JWST F323N& 32370    &   0.0334 &	0.4304   \\
JWST F444W &43441.9 & 0.0194 &	0.2504  &  JWST F405N& 40520    &   0.0219 &	0.2820   \\
JWST F150W2&15423.1 & 0.1519 &	1.9577  &  JWST F466N& 46540    &   0.0174 &	0.2242   \\
JWST F322W2&30750.5 & 0.0370 &	0.4773  &  JWST F470N& 47080    &   0.0171 &	0.2203   \\
\hline
\end{tabular}
\end{center}
\end{table}

Based on the determined NIR extinction law, 
we can predict the relative extinction in the other NIR bandpasses. 
The relative extinction values $A_\lambda/A_V$ and $A_\lambda/A_\Ks$ for the NIR bandpasses of the {\it HST} WFC3 and the {\it JWST} NIRCAM are evaluated. 
The adopted effective wavelength $\lambda_{\rm eff}$\footnote{The values of effective wavelength $\lambda_{\rm eff}$ are from the website of CMD 3.0 input form: http://stev.oapd.inaf.it/cgi-bin/cmd$\_$3.0.} and the predicted extinction results $A_\lambda/A_V$ and $A_\lambda/A_\Ks$ are tabulated in Table~\ref{tab:JWST}.  
The accuracy of this predicted NIR extinction is about 2.5\%. 
Note that the IR absorption features (Draine 2003; Fritz et al.\ 2011; Wang et al.\ 2013), 
such as ice features at $3.1\mum$ (H$_2$O), $4.27\mum$ (CO$_2$), and $4.67\mum$ (CO), 
and the aliphatic hydrocarbons feature at $3.4\mum$, 
can also affect some entries in Table~\ref{tab:JWST}. 

\section{Conclusion}

We have investigated the optical to mid-IR extinction law for a group of RC stars that were selected by the stellar parameters from the APOGEE survey. 
The multiband photometric data are collected from {\it Gaia}, APASS, SDSS, Pan-STARRS1, 2MASS, and {\it WISE} surveys. 
As the extinction tracers (RC stars) cover all the fields surveyed by APOGEE hence our measurements represent the average extinction. 
Thanks to the unprecedented {\it Gaia} data, not only a much-improved extinction law is determined, but also some issues are revealed and discussed.
The main results of this work are as follows: 

\begin{enumerate}
\item The color--excess method is adopted to derive multiband CERs $E(\lambda - \GRP)/E(\GBP - \GRP)$ for two APASS bands ($B, V$), five SDSS bands ($u, g, r, i, z$), five Pan-STARRS1 bands ($g, r, i, z, y$), three 2MASS bands ($J, H, \Ks$), three {\it WISE} bands ($W1, W2, W3$), and one {\it Gaia} band ($G$). 
We found that in the color--color or the CE--CE diagrams, the CERs display different amounts of curvature at different bands.  
This is due to the assumption of a static wavelength for each filter in determining the slopes.  
We performed curvature analysis in Section 3.3 and calibrated the curvature of CERs in the final determination of the CERs. 
Through elaborate uncertainties analysis and simulation analysis, we conclude that the total uncertainties of our CERs are less than 0.015, which validate the process of curvature correction. 
Our CERs agree with previously published values, while the precision has improved significantly. 
\item With parallaxes from {\it Gaia} DR2, the relative extinction $\ABP/\ARP=1.700\pm0.007$ is determined by the color excess--extinction method. 
This value agrees with the result from color--excess method. 
Then, we convert the CERs to the relative extinction $A_\lambda/\ARP$ with the derived $\ABP/\ARP$.
The corresponding extinction coefficients $A_\lambda/E(\GBP-\GRP)$ are also derived, including for three {\it Gaia} bands $\AG=(1.890\pm0.015)E(\GBP-\GRP)$, 
$\ABP=(2.429\pm0.015)E(\GBP-\GRP)$, and $\ARP=(1.429\pm0.015)E(\GBP-\GRP)$. 
\item A new extinction law is determined, and it can be described by $\Rv=3.16\pm0.15$ from the definition of $\Rv$. 
Comparing it with the model extinction, the CCM $\Rv=3.1$ extinction curve could only well explain this observed extinction law in the wavelength range of 300--550 nm. 
At long wavelength, to agree with the observed extinction law, an adjustment of the parameters in the CCM $\Rv=3.1$ model is made.  
It is worth mentioning that the NIR bands obey a steep extinction law, which is $A_{\lambda}/\Av = (0.3722\pm0.0026) \lambda^{-2.070\pm0.030}$. The consistency between this adjusted extinction model and the observed extinction law is better than 2.5\%. 
\item It is worth noting that the observed reddening/extinction tracks in the ISM are curved. 
The amount of curvature depends on the spectral type, the filter system, and the amount of extinction.
Hence, we also need to do the curvature correction when applying the extinction correction, particularly in the heavy extinction region.   
Otherwise, extinction laws are only safe to be used for low-reddening objects or a photometric system with extremely narrow bandwidth. 
\item
In the determination of CERs by fitting the CE--CE diagrams, the $x$-axis error (including photometric error and intrinsic color error), the sample number, and the average reddening value of high-extinction sources have great impact on the slope (CER), especially when the $x$-axis error is large.   
These are the reasons why various NIR CERs were reported in previous works. 
Compared to the $E(J-\Ks)$, the $E(\GBP-\GRP)$ is at least a factor of 3 more precise.  
Therefore, we recommend adopting high photometric precision bands (here, $\GBP$ and $\GRP$) as basis bands in the CER analysis to reduce the error of the slope caused by the fitting method. 
\item Based on the determined NIR extinction law, we predict the relative extinction values $A_\lambda/A_V$ and $A_\lambda/A_\Ks$ for the NIR bandpasses of the {\it HST} WFC3 and the {\it JWST} NIRCAM. 
\end{enumerate}


\acknowledgments{We thank Dr. Hassen Yesuf for polishing the language and Prof. Bi-Wei Jiang and Prof. Jian Gao for very helpful discussions. 
We thank the anonymous referee for very insightful and useful suggestions to improve the quality and readability of the paper. 
This work is supported by the Initiative Postdocs Support Program (No.\ BX201600002), the National Natural Science Foundation of China (No.\ U1631104).  
S.W. acknowledges support from a KIAA Fellowship. 
X.C. acknowledges support by the China Postdoctoral Science Foundation (grant 2017M610998). 

This work has made use of data from the surveys by {\it Gaia}, SDSS, Pan-STARRS1, APASS, 2MASS and {\it WISE}. 
This work has made use of data from the European Space Agency (ESA) mission {\it Gaia} (https://www.cosmos.esa.int/gaia), processed by the {\it Gaia} Data Processing and Analysis Consortium (DPAC, https://www.cosmos.esa.int/web/gaia/dpac/consortium). Funding for the DPAC has been provided by national institutions, in particular the institutions participating in the {\it Gaia} Multilateral Agreement.
This work has made use of SDSS data (http://www.sdss.org).
Funding for the Sloan Digital Sky Survey IV has been provided by the Alfred P. Sloan Foundation, the U.S. Department of Energy Office of Science, and the Participating Institutions. SDSS-IV acknowledges support and resources from the Center for High-Performance Computing at the University of Utah.
This work has made use of Pan-STARRS1 data (https://outerspace.stsci.edu/display/PANSTARRS). Fourteen organizations in six nations (plus two funding organizations) supported the Pan-STARRS1 survey. 
This research has made use of the APASS database, located at the AAVSO web site. Funding for APASS has been provided by the Robert Martin Ayers Sciences Fund.
The Two Micron All Sky Survey is a joint project of the University of Massachusetts and the Infrared Processing and Analysis Center/California Institute of Technology, funded by the NASA and the NSF. 
The Wide-field Infrared Survey Explorer is a joint project of the University of California, Los Angeles, and the Jet Propulsion Laboratory/California Institute of Technology, funded by the NASA.
}


\end{document}